\documentclass[10pt,journal,compsoc]{IEEEtran}
\ifCLASSOPTIONcompsoc
  \usepackage[nocompress]{cite}
\else
  \usepackage{cite}
\fi
\ifCLASSINFOpdf
\else
\fi
\usepackage{amsmath,amssymb,amsfonts}
\usepackage{algorithmic}
\usepackage{graphicx}
\usepackage{textcomp}
\usepackage{natbib}
\usepackage{caption}
\usepackage{array}
\usepackage[table]{xcolor}
\usepackage{colortbl}
\usepackage{tabularx}
\usepackage{multirow}
\usepackage{multicol}
\usepackage{wrapfig}
\usepackage{soul}
\usepackage{enumitem}

\usepackage{url}
\usepackage{hyperref}

\begin{document}
\captionsetup[figure, table]{font=scriptsize}

%
\title{Socio-Technical Grounded Theory \\for Software Engineering}

%
%
%



\author{Rashina Hoda
\IEEEcompsocitemizethanks{\IEEEcompsocthanksitem Rashina Hoda is with the Faculty of Information Technology, \\Monash University, Melbourne, Australia.\protect\\
E-mail: rashina.hoda@monash.edu}
\thanks{IEEE Transactions on Software Engineering (Aug 2021)}}


\markboth{Accepted for publication in IEEE Transactions on Software Engineering (Aug 2021) DOI: 10.1109/TSE.2021.3106280}{}

\IEEEtitleabstractindextext{%
\begin{abstract}
Grounded Theory (GT), a sociological research method designed to study social phenomena, is increasingly being used to investigate the human and social aspects of software engineering (SE). However, being written by and for sociologists, GT is often challenging for a majority of SE researchers to understand and apply. Additionally, SE researchers attempting ad hoc adaptations of  traditional GT guidelines for modern socio-technical (ST) contexts often struggle in the absence of clear and relevant guidelines to do so, resulting in poor quality studies. To overcome these research community challenges and leverage modern research opportunities, this paper presents \textit{Socio-Technical Grounded Theory} (STGT) designed to ease application and achieve quality outcomes. It defines what exactly is meant by an ST research context and presents the STGT guidelines that expand GT's philosophical foundations, provide increased clarity and flexibility in its methodological steps and procedures, define possible scope and contexts of application, encourage frequent reporting of a variety of interim, preliminary, and mature outcomes, and introduce nuanced evaluation guidelines for different outcomes. It is hoped that the SE research community and related ST disciplines such as computer science, data science, artificial intelligence, information systems, human computer/robot/AI interaction, human-centered emerging technologies (and increasingly other disciplines being transformed by rapid digitalisation and AI-based augmentation), will benefit from applying STGT to conduct quality research studies and systematically produce rich findings and mature theories with confidence.
\end{abstract}

}

\maketitle
\IEEEdisplaynontitleabstractindextext
\IEEEpeerreviewmaketitle

\IEEEraisesectionheading{\section{Introduction}\label{sec:introduction}}

\IEEEPARstart{G}{rounded} theory is a complete research method that enables systematic and evidence-based development of theory. The emergence of the  Grounded Theory (GT) method challenged the dominance of quantitative research in sociology in the mid-1960's and marked an emphatic resurgence of qualitative research. Since its introduction \citep{glaser1967discovery}, GT has evolved into three versions: the \textit{Classic} or \textit{Glaserian} version by one of the founding fathers, Barney Glaser, who continued to support the original method \citep{glaser1967discovery, glaser1992basics}, the second being the  \textit{Strauss-Corbinian} version introduced by the other founding father, Anselm Strauss, in collaboration with Juliet Corbin \citep{strauss1990basics}, and the third, \textit{Constructivist GT} version, introduced by Kathy Charmaz nearly two decades ago \citep{charmaz2006constructing}.

GT is a \textit{deep dive} into an area of interest or phenomenon. It is particularly suited to understudied areas, phenomena with gaps between research and practice, and for addressing complex and deep questions of \textit{how} and \textit{why} in addition to the \textit{what} of the phenomenon being studied. While GT is said to be a general method open to quantitative data, it is predominantly applied as a qualitative method through traditional data collection techniques such as interviews and observations. GT's unique features include rigorous qualitative data analysis procedures such as \textit{open coding} and \textit{constant comparison} and systematic theory development. Using an iterative, incremental, and interleaved approach to data collection and analysis, GT is one of the most agile research methods \citep{hoda2012developing}. When applied well, GT enables the researcher to undertake an up-close and in-depth exploration of a practical phenomenon and emerge with original, relevant, and parsimonious theories, distinguishing it from other qualitative methods such as case studies \citep{runeson2009guidelines}, ethnography \citep{sharp2016role}, survey research, and interview based studies.

GT continues to be popular  beyond its native field of sociology and is widely applied to study social phenomena in areas such as psychology \citep{psychGT1, psychGT2}, nursing \citep{nursingGT}, and medical education \citep{medicalGT1, medicalGT2}. In the last decade, GT has become increasingly popular in software engineering (SE) to investigate its human and social aspects. It has been applied successfully to explain SE phenomena such as \textit{software process improvement} \citep{coleman2007using}, \textit{self-organization} \citep{hoda2010organizing, hoda2012developing}, \textit{agile architecture} \citep{waterman2015much}, \textit{design problems} \citep{sousa2018identifying}, and \textit{self-assignment} \citep{masood2020emse}. Table \ref{tab:ExemplarGTstudies} presents a list of exemplar GT in SE studies.

However, a majority of SE researchers struggle with applying GT, as evident from a critical review of GT in SE studies over 20 years which identified major quality concerns such as studies not acknowledging the version of GT being applied, combining guidelines without rationales, and applying GT practices \textit{\`{a} la carte} with no reference to particular guidelines \citep{stol2016grounded}. The review reported other concerns such as only 15\% studies provided detailed accounts of their research procedures and fewer made theoretical contributions. Such poor quality show is indicative of researcher apprehension, misunderstanding, poor presentations, and extreme evaluations, in addition to possible method misuse and abuse. 

And yet, these findings are not surprising. Glaser and Strauss would have hardly had the SE researcher in mind when introducing GT. The \textit{traditional GT guidelines} are spread across the fundamental texts of the three GT versions \citep{glaser1967discovery, glaser1978theoretical, strauss1990basics, charmaz2006constructing, charmaz2014constructing}. They declare theory development to be exclusively interesting to and achievable by ``\textit{only sociologists [who] are trained to want it, look for it, and to generate it}'' \citep{glaser1967discovery, glaser2017discovery}. They were written by and for sociologists using language, format, and examples native to them, making them difficult for SE researchers to easily access and understand. At the same time, decontextualising GT is acknowledged to lead to n\"aive applications \citep{martin2019GT}, pointing to the need for contextualised GT guidelines for SE research. While exemplar studies (Table \ref{tab:ExemplarGTstudies}) and dedicated papers describing GT guidelines for SE research provide good starting points \citep{hoda2012developing, adolph2011using, stol2016grounded}, they do not provide enough independent guidance for novice SE researchers on adapting and applying GT in SE contexts. Currently, achieving high quality applications and outcomes relies on first understanding the traditional GT guidelines and then effectively selecting and applying one of the three versions.

GT is acknowledged to be ``\textit{fundamentally aimed at explaining and rendering convincing portrayals of social processes}'' \citep{timonen2018GTchallenges}. Software engineering, on the other hand, is a fundamentally socio-technical domain \citep{whitworth2011social, storey2020EMSE}. SE abounds with phenomena that are neither exclusively social nor purely technical, rather, predominantly \textit{socio-technical} (ST) where \textit{the social and technical aspects are interwoven in a way that studying one without due consideration of the other makes for an incomplete investigation and understanding}. Requirements elicitation, user-centred design, collective estimations, pair programming, daily standups, multi-team coordination, acceptance testing, maintenance, and DevOps are examples of ST phenomena, where a complete research investigation needs to consider the full ST context. Using GT's social science approach to studying socio-technical phenomena in SE will only go so far before adaptations and updates are required.

Indeed the most successful GT in SE studies have been those that have overcome the entry barriers of understanding traditional GT guidelines and applied them, often with implicit adaptations, to work in ST contexts. However, as evident from the GT in SE quality review, they are exceptions not the norm \citep{stol2016grounded}. How can SE researchers, a majority of whom struggle with basic understanding of traditional GT guidelines, be expected to select one of three versions and adapt and apply them effectively without any explicit philosophical and methodological guidance? 

Critically, there are no comprehensive guidelines available for SE researchers to adapt and apply GT simply and effectively in SE's modern ST research contexts. Reviewer proclamations of ``\textit{this is not GT!}'' and recommendations to retrospectively rebrand attempted GT applications and adaptations as interview-based or mixed methods studies do not fix the underlying misalignment and guidance gap.

Finally, SE is the birthplace of modern data collection and analysis tools and techniques that are revolutionising research. For example, publicly available data on social media \citep{storey2010socialmedia} and open source repositories, and modern data collection and analysis techniques such as mining software repositories, natural language processing including sentiment analysis \citep{novielli2018sentiment}, and artificial intelligence based tools. While traditional GT guidelines do not explicitly overrule such advancements, they obviously do not address them, leaving individual SE researchers struggling with ad hoc adaptations \citep{stol2016grounded}. SE is particularly well positioned to leverage modern software-driven tools and techniques to enhance GT practice within and beyond SE research, pointing to the need for a contemporary adaptation and update of the traditional GT guidelines so practitioners can systematically apply them.

Adaptations and updates to traditional research methods from time to time are not uncommon. For example, contextualised guidelines for other sociological research methods such as \textit{Case Studies} \citep{runeson2009guidelines} and \textit{Ethnography} \citep{sharp2016role} have made it easier for SE researchers to apply and succeed with these methods. Modern updates to traditional methods, such as \textit{Virtual ethnography} \citep{hine2008virtual} and \textit{Visual ethnography} \citep{schwartz1989visual, pink2020doing} have provided the necessary guidance to harness modern technological advancements while leveraging the best of traditional methods. \textit{Constructivist GT} provided a constructivist update to traditional GT guidelines \citep{charmaz2006constructing}.

Similarly, motivated by the needs of, and opportunities afforded by, the SE research community, I present \textit{Socio-Technical Grounded Theory} that aims to ease accessibility and improve GT quality outcomes in SE by:
\begin{itemize}
    \item defining the socio-technical research context as it applies to SE research,
    \item presenting simple and contextualised guidelines for conducting and evaluating STGT research,
    \item explicitly listing the fundamental knowledge and skills required to achieve quality outcomes,
    \item expanding its philosophical foundations to accommodate a range of paradigms and perspectives,
    \item offering the option to apply the \textit{full STGT method} or \textit{STGT for Data Analysis} within other methods,
    \item offering guidelines on harnessing modern data types, sources, collection tools, and analysis techniques,
    \item offering the option to select from emergent or structured theory development modes later in the process, once experience with basic data collection and analysis is gained,
    \item encouraging frequent reporting of a variety of interim, preliminary, and mature outcomes,
    \item presenting nuanced evaluation guidelines for different types of applications and outcomes.
\end{itemize}

As with any guidelines, practical applications will serve to provide feedback for further improvements.


\section{Grounded Theory -- Social Traditions}

\subsection{The Origins}
Grounded Theory (GT) emerged in the sociological research scene at a time when quantitative research with its focus on objectivity, accuracy, verification, and generalisability had taken a strong foothold following similar trends in the wider research communities of the natural sciences disciplines. Qualitative research, on the other hand, was being seen as lacking scientific rigour and relying on anecdotal evidence. With its systematic and rigorous techniques and procedures, the introduction of GT challenged the status quo and served to re-establish the value of qualitative research in sociology. It challenged the overemphasis on \textit{theory verification} predominant at the time and argued for \textit{theory development} as a fundamental research objective.

GT originated from the research of sociologists, Barney G. Glaser and Anselm L. Strauss, as they studied the awareness of dying patients and the role of their social status in the level of nursing care received, interspersed across three texts, \textit{Awareness of Dying} \citep{glaser1966awareness}, \textit{Time for Dying} \citep{glaser1980time}, and \textit{Status Passage} \citep{glaser2011status}. The procedures and strategies employed in their qualitative research project led to the formulation of the Grounded Theory method, documented into their book ``\textit{The Discovery of Grounded Theory -- strategies for qualitative research}'' \citep{glaser1967discovery}.

\subsection{The Method}
Grounded theory is a rigorous research method that enables systematic and evidence-based development of theory. GT is a complete research method, covering data collection, analysis, and more advanced steps of theory development. GT enables the investigation of social phenomena using a social lens native to sociology. The name grounded theory is derived from its focus on \textit{theory} development firmly \textit{grounded} in evidence collected from real-world practice. 

Distinguishing features of GT include its unique focus on inductive theory development through \textit{iterative} and \textit{interleaved} rounds of data collection and analysis. While literature review is not strictly forbidden in GT, it is encouraged in "unrelated fields" so as not to bias theory development in the area being studied (pg35, \cite{glaser1992basics}.) A GT study begins with evidence collection from the substantive field, typically through semi-structured interviews and observations. GT offers a set of data analysis procedures such as \textit{open coding}, \textit{constant comparison}, and \textit{selective coding} that lead to the inductive identification of \textit{key patterns}, referred to as \textit{concepts} and \textit{categories}. Rigorous application of the GT method leads to the development of theories which encapsulate the key patterns and relationships between them.

Theory development in GT is facilitated by researcher traits such as \textit{theoretical sensitivity} \citep{glaser1978theoretical, strauss1990basics} and procedures such as \textit{memoing} and \textit{theoretical- sampling}, \textit{sorting}, \textit{coding}, and \textit{saturation}. Theoretical sensitivity is the propensity of the researcher to develop theoretical codes, structures, relationships, and eventually theories. Unlike field and observational notes that are focused on documenting full accounts of observed practice in a single instance, \textit{memoing} enables the researcher to explore emerging concepts, potential relationships between concepts compared across instances, and identify gaps in the emerging theoretical structures. 

Guided by the theoretical gaps, researchers can decide where and from whom to collect data in the next collection-analysis cycle and refine collection procedures such as interview questions accordingly, a process known as \textit{theoretical sampling}. \textit{Theoretical sorting} involves examining, arranging, and rearranging theoretical memos to enable story lines or theoretical structures to emerge. These emerging structures can be presented using best matching predefined theory templates, an optional procedure called \textit{theoretical coding}. Once collecting more data reaches a point of diminishing returns, the study is said to have reached \textit{theoretical saturation}. These traditional GT strategies, techniques, and procedures are described at length in the traditional texts \citep{glaser1967discovery, glaser1978theoretical, glaser1998doing, strauss1990basics, strauss1994grounded, charmaz2006constructing, charmaz2014constructing}.

The resulting theory derived from the application of GT is said to be a \textit{mid-range} theory, limited in its applicability to the contexts studied \citep{glaser1967discovery}.  However, it remains open to future modifications in light of new evidences gained from further investigations.

\begin{table*}[t]
\scriptsize
\caption{\small Some exemplar GT studies in software engineering presenting theories and providing comprehensive and detailed documentation of GT application (*distinguished paper award)}
\vspace{-0.5cm}
\label{tab:ExemplarGTstudies}
\begin{tabular}{>{\raggedright\arraybackslash}p{3.4cm}>{\raggedright\arraybackslash}p{14cm}}\\
\noalign{\smallskip}\hline\noalign{\smallskip}
\cite{coleman2007using} & \textit{Using grounded theory to understand software process improvement...} Information and Software Technology. \\
\cite{hoda2010organizing} & \textit{Organizing self-organizing teams}, IEEE International Conference on Software Engineering.\\
\cite{dagenais2010moving} & \textit{Moving into a new software project landscape}, IEEE International Conference on Software Engineering.\\
\cite{hoda2011impact} & \textit{The impact of inadequate customer collaboration on self-organizing Agile teams}, Information and Software Technology.\\
\cite{hoda2012self} & \textit{Self-organizing roles on agile software development teams}, IEEE Transactions on Software Engineering.\\
\cite{adolph2012reconciling} & \textit{Reconciling perspectives: A grounded theory of how people manage the process of software development}, Journal of Systems and Software.\\
\cite{maglyas2013roles} & \textit{What are the roles of software product managers? An empirical investigation}, Journal of Systems and Software.\\
\cite{jantunen2014using} & \textit{Using a grounded theory approach for exploring software product management challenges}, Journal of Systems and Software.\\
$^*$\cite{waterman2015much} & \textit{How much up-front? A grounded theory of agile architecture}, IEEE International Conference on Software Engineering.\\
\cite{stray2016daily} & \textit{The daily stand-up meeting: A grounded theory study}, Journal of Systems and Software.\\
$^*$\cite{hoda2017becoming} & \textit{Becoming agile: A grounded theory of agile transitions in practice}, IEEE International Conference on Software Engineering.\\
\cite{sedano2017software} & \textit{Software development waste}, IEEE International Conference on Software Engineering.\\
$^*$\cite{sousa2018identifying} & \textit{Identifying design problems in the source code: A grounded theory}, IEEE International Conference on Software Engineering.\\
\cite{masood2020emse} & \textit{How agile teams make self-assignment work: a grounded theory study}, Empirical Software Engineering.\\
\cite{masood2020tse} & \textit{Real world scrum a grounded theory of variations in practice}, IEEE Transactions on Software Engineering.\\
\cite{yogi2021jss} & \textit{The role of the project manager in agile software development projects}, Journal of Systems and Software.\\
\cite{yogi2021scrummaster} & \textit{Spearheading agile: the role of the scrum master in agile projects}, Empirical Software Engineering.\\
\noalign{\smallskip}\hline\noalign{\smallskip}
\end{tabular}
\end{table*}

\subsection{The Evolution}
\label{GTEvolution}
Since its introduction in 1967, the GT method has evolved -- been adapted and has changed forms over the years -- through the introduction of two other versions. Nearly two decades later, one of the two founding fathers, Anselm Strauss along with Juliet Corbin proposed the first variation of GT in their book ``\textit{Basics of Qualitative Research}'', which came to be known as  \textit{\textbf{Strauss-Corbinian GT}} \citep{strauss1990basics}. The original GT method supported by Glaser is since referred to as \textit{\textbf{Classic}} or \textit{\textbf{Glaserian GT}}. Kathy Charmaz introduced a third version with her book ``\textit{Constructing Grounded Theory}'' \citep{charmaz2006constructing} a decade and a half later.

While Glaser maintained that theory emerges naturally from the underlying data \citep{glaser1978theoretical}, Strauss and Corbin introduced a more prescriptive and structured way to derive theory \citep{strauss1990basics}. A distinguishing feature of the Strauss-Corbinian approach was the introduction of \textit{axial coding}, a ``\textit{process of systematically relating categories and sub-categories}''. For this, they recommended the use of analytical tools such as the \textit{coding paradigm}, a sort of template structure of a theory with consideration of the context, causal conditions, intervening conditions, action-interaction strategies, and consequences related to the phenomenon. Using such tools (e.g. coding paradigm or conditional matrix) is likely to result in the theory being structured sooner in the study as compared to \textit{Classic GT}, and verified through deductive approaches in the later parts of the study. Its structured approach is a reason why some researchers prefer  \textit{Strauss-Corbinian GT} \citep{masood2020emse, masood2020tse}.

However, Glaser strongly objected to this variation as ``\textit{whole different method}'' and the resulting theory as ``\textit{forced}'' rather than emergent, captured in his rejoinder ``\textit{Basics of Grounded Theory Analysis: Emergence vs. Forcing}'' \citep{glaser1992basics}. Meanwhile, with the passing away of Strauss in 1996, Corbin continued to publish their co-written second edition and a  third edition while maintaining their joint position.

The next chapter in the evolution of GT was marked by a third version formulated by Kathy Charmaz in her book ``\textit{Constructing Grounded Theory}'' \citep{charmaz2006constructing}. Charmaz described the research paradigms underlying Glaserian and Strauss-Corbinian GT as primarily \textit{positivist} and the latter with some \textit{pragmatic} influences. Charmaz proposed \textit{\textbf{Constructivist GT}}, a constructivist approach to GT, highlighting the role of researcher in subjectively interpreting and \textit{constructing} reality as opposed to an objectivist stance where the researcher is seen as neutral, objectively observing and rendering reality \citep{charmaz2014constructing}. Charmaz's plenary presentation in 1993 ``\textit{caused quite a stir}'' with opinions prominently divided along gender lines. Though initially supported mainly by women \citep{charmaz2014constructing}, Constructivist GT subsequently had many takers, becoming a prominent version of GT. Dedicated articles discussing the differences between the versions in-depth can be consulted in \citep{GTDifferences2, GTDifferences1, kenny2014tracing}. 

\section{GT in Software Engineering}
\subsection{The Promise}
\label{GTsweetspot}
Theories have the potential to be useful to both research and industry \citep{sjoberg2008theory}. SE researchers interested in theory development will benefit from using GT, which promises theory development as its core feature. GT can help researchers lay down the theoretical foundations of SE, the youngest engineering discipline \citep{50yearsSE}, explaining its unique landscape, practices, challenges, and strategies.

Being an empirical method, GT promises combining research rigour with practical relevance -- described as a '\textit{grand challenge}' \citep{gregory2016challenges}. Theories developed using GT have the potential to lay the foundations of better understanding SE phenomena, present recommendations and guidelines, and motivate tools development. For example, a GT study on \textit{the role of ethics in artificial intelligence} can lead to theories that explain how developers perceive ethics, the socio-technical barriers to embedding ethics in AI, and enabling strategies. Practical applications can then be drawn in terms of practitioner guidelines for embedding ethics in AI. Similarly, grounded theories can establish rich theoretical foundations of the problem space, leading to development of software tools and techniques.

Like most methods, GT has a \textit{sweet spot} -- an ideal context where it is most likely to succeed and indeed where it makes most sense to use. The application of GT promises to benefit SE studies that fall within the traditional GT sweet spot.\\


\begin{itemize}

\item \textbf{Human and social aspects focus}. Originating from sociology, it is no surprise that GT is aptly suited to studying human and social aspects of SE.

\item \textbf{Theory development}. GT is particularly suited to studying areas lacking existing theories, those with research gaps, and where existing research fails to resonate with practice and can benefit from new theories grounded in empirical evidence.

\item \textbf{Practice-based topics}. If the research topic is practice-based or industry relevant, it is more likely to lead to a successful GT because it increases the possibility of finding enough evidence from practice and for the researcher to conduct field observations.

\item \textbf{Complex and deep questions}. While GT can be employed to answer the \textit{what} type questions, it is best used to explore and answer more complex and deep issues through the \textit{how} and \textit{why} type questions.

\item \textbf{Primarily qualitative studies}. While GT is described as a general research method \citep{glaser1967discovery} capable of incorporating both qualitative and quantitative data, there is limited guidance on achieving this. Practically, GT is most often applied as a qualitative research method due to its rigorous qualitative data analysis procedures and results in what \cite{storey2020EMSE} refer to as \textit{descriptive} knowledge using their design science approach to categorising research outcomes.

\item \textbf{Data collected through interviews and observations.} While GT is open to including all types of data, interviews and observations form the most commonly used data sources in a traditional GT study.
\end{itemize}

On the other hand, research contexts relying primarily on quantitative data, exploring simple questions that can be answered using descriptive analysis, and those not interested in deriving theoretical frameworks, theories, or models, do not require a GT approach. Unsurprisingly, a majority of GT studies in software engineering in the last decade fall within the traditional GT sweet spot.

\begin{table*}[]
    \centering
    \caption{\small Traditional GT context and limitations, associated SE challenges presented as \#patterns, and STGT advantages.}
    \scriptsize
    \begin{tabular}{>{\raggedright\arraybackslash}p{6.8cm}>{\raggedright\arraybackslash}p{3cm}>{\raggedright\arraybackslash}p{7cm}}
    \noalign{\smallskip}\hline\noalign{\smallskip}
    \textbf{Traditional GT Context and Limitations} & \textbf{Associated SE Challenges} & \textbf{Socio-Technical GT Advantages} \\
    \hline\noalign{\smallskip}
    Traditional GT guidelines were written by and for sociologists. & \#EagerButNotEquipped & STGT method is written by and for ST researchers using format, language, and examples native to SE. \\
    Traditional GT guidelines are spread across multiple books and three  versions: \textit{Glaserian}, \textit{Strauss-Corbinian}, \textit{Constructivist}. & \#NoVersionControl & STGT guidelines present one research method, offering flexibility in its application.\\
    Traditional GT guidelines expect the researcher to understand all three versions and select one upfront. & \#NoVersionControl & STGT delays the decision about theory development modes, \textit{emergent} or \textit{structured}, after some experience is gained.\\
    Traditional GT guidelines assume that the researcher is aware of and possesses the requisite knowledge and skills for conducting successful GT studies. & \#EagerButNotEquipped \#PoorPresentations \#ExtremeReviews & STGT guidelines explicitly spell out the fundamental knowledge and skills for conducting successful STGT studies.\\
    Traditional GT guidelines focus on theory generation. & \#ScopeConfusion \#ExtremeReviews & A \textit{full STGT study} can be used to generate theories and \textit{STGT for Data Analysis} can be used for a data analysis only within other research frameworks, e.g. case study.\\
    Traditional GT methods were designed to study social phenomena. & \#DIYGoneWrong \#ExtremeReviews & STGT method enables the investigation of socio-technical phenomena and domains.\\
    Traditional GT guidelines were designed to use traditional data types, sources, and collection techniques. & \#DIYGoneWrong \#ExtremeReviews & STGT method is designed to leverage traditional and modern data types, sources, and collection techniques.\\
    Glaserian and Strauss-Corbinian GT were designed to be standalone methods. & \#DIYGoneWrong \#ExtremeReviews & STGT method can be applied standalone or in combination with other methods and techniques.\\
    Traditional GT guidelines did not have to consider the need to share detailed evidence. & \#EvidenceGateKeeping \#TheFacade & STGT method provides guidelines on presenting sufficient sanitised evidence and examples to establish credibility and enable proper evaluation.\\
    \noalign{\smallskip}\hline
    \end{tabular}
    \label{tab:TraditionalGTLimits}
\end{table*}

\subsection{The Progress}
Both the SE discipline and the GT research method are about half a century old and still maturing in practice. The two have continued to progress in parallel with their paths crossing over as SE researchers discovered and attempted GT. The use of GT has gained steady acceptance in SE research, particularly in the last decade \citep{stol2016grounded}.

The growing appeal of GT in SE research is being propelled on the one hand by \textit{theory development efforts} \citep{stol2016grounded} and on the other, by the \textit{rise of human and social aspects} in SE \citep{prikladnicki2013CHASE} and \textit{agile software development} \citep{hoda2018rise}. GT has been used to study a variety of human and social aspects of SE. Some exemplar GT studies in software engineering that describe both their method application and the resulting theory in sufficient detail are listed in Table \ref{tab:ExemplarGTstudies}. Dedicated SE papers providing guidance on traditional GT methods include \cite{adolph2011using}, \cite{hoda2012developing}, and \cite{stol2016grounded}.

\subsection{The Challenges}
\label{GTChallenges}
A critical review of GT in SE studies identified increasing adoption over the last two decades but raised serious quality concerns \citep{stol2016grounded}. Of the final 98 journal papers analysed, half explicitly claimed to perform GT while only a third provided details of their method application. These were seen to apply guidelines primarily from the \textit{Glaserian} or \textit{Strauss-Corbinian} versions. A majority of studies did not acknowledge the version of GT being employed. Some combined guidelines from different versions without rationale. Several were seen to apply practices \textit{\`{a} la carte} with no reference to any guidelines. Only five individual studies were found to be comprehensive and detailed in their GT application description, serving as exemplars \citep{coleman2007using}, \citep{hoda2011impact}, \citep{adolph2012reconciling}, \citep{hoda2012self}, and \citep{jantunen2014using}.

Over fifteen years of experience in conducting, supervising, reviewing, and editing GT in SE studies suggests these widespread issues are not always intentional method misuse or abuse. \textit{Method slurring}, false claims of applying method guidelines \citep{methodSlurring}, is but one reason quality issues arise. Many genuine attempts fail or lead to poor results and harsh reviews, indicating researcher apprehensions, misunderstandings, poor presentations, and extreme evaluations underlying the poor show of quality, captured in the following patterns of GT in SE challenges.\\

\noindent \textbf{\#EagerButNotEquipped -- \textit{When software engineering researchers are interested in using GT but struggle to understand and apply it}}. The traditional GT texts are written by and for sociologists and are naturally challenging for SE researchers, many of whom have little or no sociological or qualitative research background. Often the traditional texts start to make sense after  preliminary data collection and analysis, as researchers connect the  guidelines with their lived experiences. While the exemplar GT studies in SE provide application details (Table \ref{tab:ExemplarGTstudies}), they are not complete methodological guidelines for the novice researcher. For the vast majority of SE researchers, GT remains largely inaccessible and challenging. Not all SE researchers need to or may want to develop theories, but for those who are interested, GT should be accessible.\\
    
\noindent \textbf{\#NoVersionControl -- \textit{When software engineering researchers are unaware or unclear about the different versions of traditional GT and how to apply them in practice}}. Despite in-depth attempts to delineate between the different versions of GT \citep{GTDifferences2, GTDifferences1, kenny2014tracing}, researchers continue to apply traditional GT versions poorly \citep{stol2016grounded}. For the novice GT researcher, understanding any one version is challenging enough. Expecting an understanding of all three, the differences between them, and selecting one upfront, further raises the entry barrier.\\
    
\noindent \textbf{\#ScopeConfusion -- \textit{When software engineering researchers find it difficult to distinguish between conducting a full GT study and using GT data analysis techniques alone}}. GT's data analysis procedures of \textit{open coding}, \textit{constant comparison}, and \textit{selective coding} are popular in qualitative research. Sometimes SE researchers use these GT data analysis techniques within some other research framework and present it in a way that claims a full GT study. While the other key GT procedures of \textit{iterative and interleaved} data collection and analysis, \textit{memoing}, \textit{theoretical sampling, sorting, coding}, and \textit{theoretical saturation} are clearly missing.\\

\noindent \textbf{\#EvidenceGateKeeping -- \textit{When software engineering researchers find it difficult to support their GT findings with adequate evidence without compromising ethics}}. There is an increasing push for open science and data transparency in the SE research community. Typically, the raw data underlying a GT study, such as interviews and observational field notes, cannot be shared outside of the core research group because of the confidentiality and anonymity clauses of the governing human ethics. Many GT studies err on the side of providing no evidence of raw data and analysis, leading to credibility concerns.\\

\noindent \textbf{\#TheFacade -- \textit{When software engineering researchers do not apply GT carefully but claim to do so}}. Sometimes, SE researchers claim to apply GT but their methodology and outcomes descriptions suggest poor and rushed applications. The increasing pressure to publish, both system permeated (e.g. tenure or promotion driven) and self-inflicted (e.g. peer pressure, international rankings, quasi-gamification of researcher profiles), manifests as increasing trends toward achieving publication quantity over quality. While not exclusive to GT, such trends encourage shallow research studies focusing on simple problem spaces, \textit{low hanging fruits}, instead of addressing complex problems through deep and meaningful work that necessitate careful method application. So long as the research community does not acknowledge and reverse these trends, method misuse and abuse will likely continue.\\

\noindent \textbf{\#PoorPresentation -- \textit{When software engineering researchers struggle to present GT studies to a high standard}}. In contrast to \#TheFacade, this pattern captures scenarios where the method was well applied but the outcome is poorly presented. Tedious presentation of the findings reading like \textit{endless walls of text} and lack of clear summaries of the theoretical and practical contributions, pertinent quotes, and evaluation make for poor presentations.\\

\noindent \textbf{\#DIYGoneWrong -- \textit{When software engineering researchers attempt to adapt traditional GT but lack guidelines to do so}}. SE researchers often attempt to combine steps and procedures across GT versions. Such ad hoc variations are often poorly executed. While some of these are rightly labelled \textit{method slurring}, other well-intended adaptations are reviewed harshly because of the absence of philosophical and methodological guidelines for adaptations.\\

\noindent \textbf{\#ExtremeReviews --  \textit{When software engineering reviewers are extreme, overly trusting or harsh, in their review of GT studies.}} Reviewers can be overly trusting based on some \textit{checklist} that authors can demonstrate having satisfied and a seemingly \textit{reasonable looking} theory. Passing poorly done GT studies as acceptable does not help improve quality. More often, reviewers are overly harsh in their evaluations because they do not fully understand GT, evaluate it using criteria fit for quantitative methods and positivist research paradigms (e.g. reproducibility, replicability), are unconvinced about the need for adaptations, and expect textbook applications of traditional GT methods using a \textit{checklist} approach. Debunking attempts at applying GT as ``\textit{this is not GT}'' or recommending a post facto re-branding to case study or interview-based study is not constructive and detrimental to future GT attempts.

\subsection{Misalignment and Need for Evolution}

\begin{quote}
    \textit{``You can't use an old map to explore a new world'' -- Albert Einstein}
\end{quote}

What makes successful GT practice so challenging in SE? \textbf{At the heart of these challenges lies the fundamental misalignment of traditional, sociological GT guidelines to study software engineering's socio-technical research context.}

While traditional GT methods have served well to study the human and social aspects of SE, successful GT studies continue to be  exceptions, not the norm in SE research \citep{stol2016grounded}. Designed in the mid 1960's by and for sociologists to study social phenomena, understandably, traditional GT methods are not well equipped to fully address SE's socio-technical research context. As a first step to adapting GT for SE, the key limitations of traditional GT methods and the new opportunities afforded by SE need to be acknowledged. For example, traditional GT guidelines are spread across multiple books and three different GT versions, leading to the \#NoVersionControl challenge. SE researchers will benefit from having a single GT method, offering flexibility in its application. Similarly, SE researchers are keen to explore modern research techniques such as sentiment analysis within GT, with some success \citep{madampe2020towards}, but it can easily lead to \#DIYGoneWrong. SE researchers will benefit from modern GT guidelines that explain the use of modern data types, sources, collection, and analysis techniques. Table \ref{tab:TraditionalGTLimits} provides a set of such contextual limitations of traditional GT guidelines, their mapping to the associated SE challenge patterns, and how these are addressed in the STGT guidelines (described in the later sections).

Traditional social science research methods have been successfully `\textit{translated}' and explained for the SE research community. \cite{runeson2009guidelines} provided guidelines for conducting and reporting \textit{Case Studies} in SE research, while \cite{sharp2016role} explained the role of \textit{Ethnography} for SE contexts, making it easier for SE researchers to apply social science methods to studying SE phenomena. Other efforts, outside of SE, to \textit{update} traditional social science methods for modern times include \textit{Virtual ethnography}, a modern update to traditional ethnography to cater to ``\textit{novel social spaces}'' \citep{hine2008virtual} and \textit{Visual ethnography} \citep{schwartz1989visual, pink2020doing}.

Similarly, a modern, contextualised update to the traditional GT guidelines will enable SE researchers to conduct socio-technical GT research with confidence, harnessing the best of traditional and modern research opportunities, tools, and techniques. The first step in this direction is to define the \textit{new world} -- the socio-technical research context (section 4), followed by describing the \textit{new map} -- the socio-technical grounded theory (STGT) method (sections \ref{method} to \ref{application}).

\section{Socio-Technical Grounded Theory}
\vspace{0.25cm}
\noindent\fbox{%
    \parbox{0.47\textwidth}{
        \textbf{Socio-Technical Grounded Theory} (STGT) is an iterative and incremental research method for conducting socio-technical research using traditional and modern research techniques to generate novel, useful, parsimonious, and modifiable theories.
        
        \textbf{Distinguishing features:} interleaved rounds of \textit{basic} data collection and analysis, \textit{emergent} or \textit{structured} mode of theory development through \textit{advanced} data collection, analysis, and theory development procedures, using primarily \textit{inductive} but also \textit{deductive} and \textit{abductive} reasoning. 
    }
}
\vspace{0.25cm}

Motivated by the common challenges of the software engineering research community in understanding and practicing traditional GT guidelines (described in section \ref{GTChallenges}) and with the aim to update traditional GT guidelines for conducting socio-technical research (defined next, in section \ref{context}), I present Socio-Technical Grounded Theory. 

STGT is an iterative and incremental research method for conducting socio-technical research using traditional and modern research techniques. It is \textit{iterative} because it involves interleaved cycles of data collection and analysis that inform and support each other. It is \textit{incremental} because each cycle progresses the research project toward  theoretical outcomes.

The STGT guidelines include its socio-technical research context (section \ref{context}), philosophical foundations (section \ref{philosophy}), methodological steps and procedures (section \ref{method} and \ref{advancedtheorydev}), its application, outcomes, and reporting (section \ref{application}), and evaluation (section \ref{evaluation}). Table \ref{tab:comparison} presents an overview of the STGT guidelines while comparing it to traditional GT guidelines.

\subsection{Socio-Technical Research}
\label{context}

\begin{figure*}[t]
    \centering
    \includegraphics[width=0.99\textwidth]{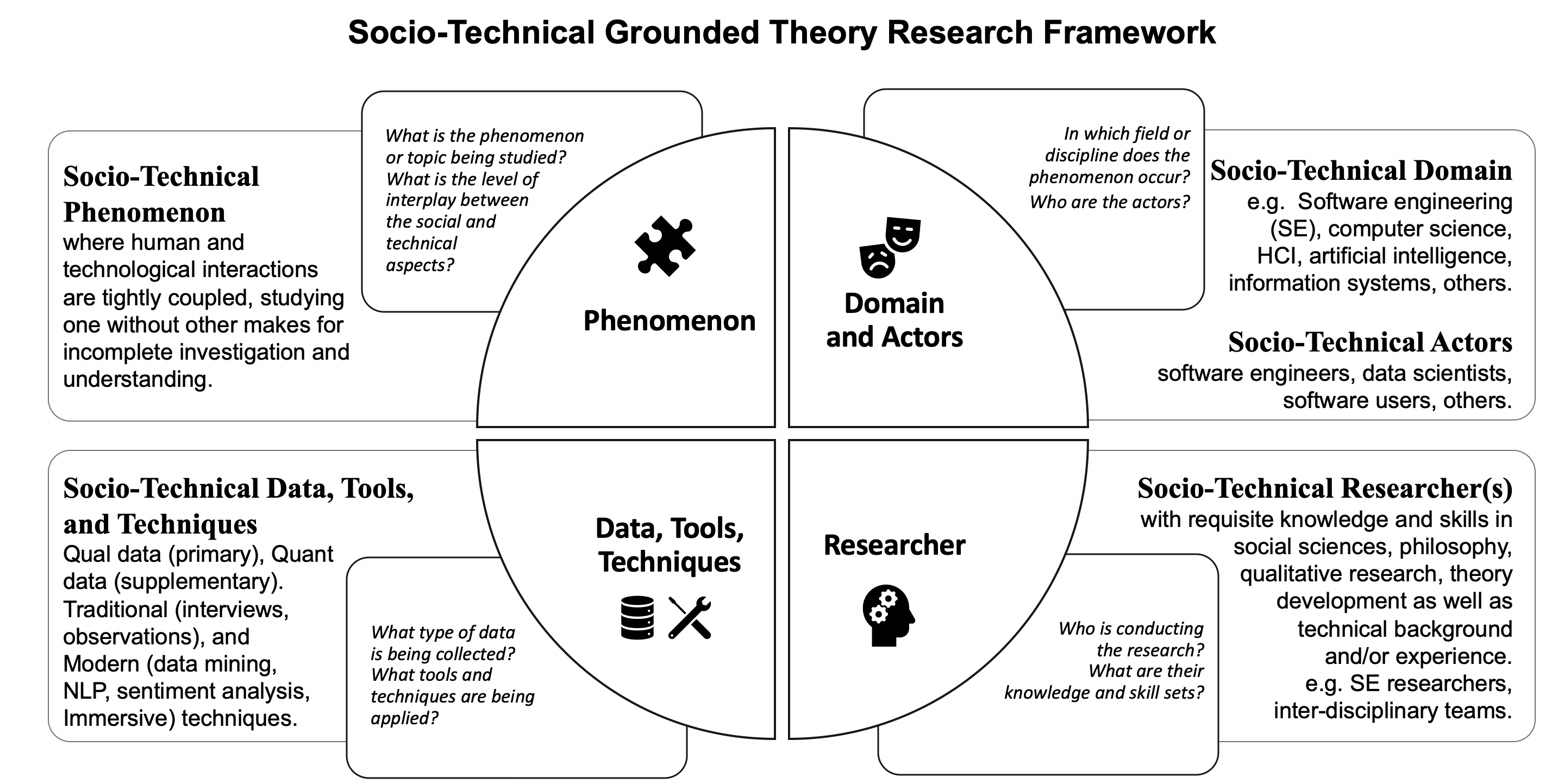}
    \caption{\small A Framework for Socio-Technical Grounded Theory Research -- socio-technical phenomenon (top left), domain and actors (top right), researchers (bottom right), and data, tools, and techniques (bottom left).}
    \label{fig:STResearch}
\end{figure*}

To understand the underlying foundations of STGT, it is important to first define what is meant by socio-technical research. Figure \ref{fig:STResearch} shows the socio-technical research framework for grounded theory, with its four dimensions:

\begin{itemize}
    \item \textbf{Phenomenon} -- \textit{what is being studied? what is the level of interplay between the social and technical aspects?}
    \item \textbf{Domain and actors} -- \textit{in which field or discipline does the phenomenon occur? who are the actors?}
    \item \textbf{Researcher} -- \textit{who is conducting the research? what are their knowledge and skill sets?}
    \item \textbf{Data, tools, and techniques} -- \textit{what is the nature of data being collected? what tools and techniques are being used?}
\end{itemize}

These dimensions capture a research landscape or context that is fundamentally different to the native contexts for which traditional GT methods were originally designed.

\subsubsection*{Socio-Technical Phenomenon}

Software engineering abounds with \textbf{socio-technical phenomenon}, \textit{where human and technological interactions are tightly coupled, such that studying one without the other makes for an incomplete investigation and understanding}. For example, the core SE practice of programming is not strictly technical, rather, it is socio-technical, involving intensive human-human collaboration and coordination and human-technology interactions. Similarly, most SE practices such as pair programming, designing prototypes, customer collaboration, software deployment, maintenance and DevOps, are all socio-technical in nature.

Research communities such as human-computer interaction (HCI), computer-supported cooperative work (CSCW), and cooperative and human aspects of software engineering (CHASE) have been at the forefront of exploring the human and social aspects. Most topics covered by these areas involve socio-technical, rather than strictly social, phenomena. More recently, with the rise of artificial intelligence (AI), there is renewed interest in gaining in-depth understanding of human factors \citep{hidellaarachchi2021effects}, human values \citep{perera2020study, hussain2020human}, and the human-in-the-loop in order to build more realistic, responsible, and usable AI-enabled systems.

Some contemporary phenomena and future topics that are particularly suited for STGT studies in SE/AI and other socio-technical domains include:

\begin{itemize}
    \item \textit{understanding human aspects (e.g. personality, gender, age, emotions, motivation, etc.) in SE/AI},  
    \item \textit{understanding human values (e.g. ethics, privacy, safety, security, morality, etc.) in SE/AI}, 
    \item \textit{understanding social aspects (e.g. collaboration, coordination, formation etc.) in SE/AI}, 
    \item \textit{role of technology (e.g. social media, virtual/remote collaboration tools) in crisis management, remote work, online education, political campaigns, gaming, etc.},
    \item \textit{human- and social-centered design of inclusive and responsible SE/AI systems},
    \item \textit{understanding socio-technical aspects of specific SE practices, e.g. requirements elicitation, user-centered design, estimation, pair/mob programming, etc}.
\end{itemize}

\noindent Some meta-research topics that will benefit from STGT studies include:
\begin{itemize}
    \item \textit{nature of human reality in the digital/AI age (ontology)},
    \item \textit{role of researcher in studying ST research (epistemology)},
    \item \textit{ethics in SE research},
    \item  \textit{credibility in SE research},
    \item  \textit{practical/industrial relevance in SE research},
    \item  \textit{bias in SE reviewing (e.g. applying inappropriate evaluation criteria: quantitative vs qualitative, research paradigm)},
    \item  \textit{enabling interdisciplinary research teams}
\end{itemize}

 Acknowledging these phenomena as socio-technical, in SE and other domains, is the first step to enabling more complete and comprehensive interpretations, constructions, and rendering of modern human realities.

\subsubsection*{Socio-Technical Domain and Actors}
With increasing proliferation of technology and digitalisation, most domains are becoming socio-technical to varying degrees, e.g. banking, medicine, education, and retail. Domains such as SE, computer science, human computer interaction (HCI), artificial intelligence, and information systems, are inherently and inextricably \textbf{socio-technical domains}, \textit{with tight coupling between its social and technical aspects}. Domains such as SE do not simply represent a user base of technology, they are the birthing cradle of information technology and systems. Technology is not only a prominent enabler or feature in these domains, it is the core raison d'être for organisations, e.g. for software companies. 

\textbf{Actors}, \textit{people who play key roles in the phenomenon}, in these domains are not regular users of technology. For example, software engineers, data scientists, computer scientists -- the actors in the SE domain -- are the creators and maintainers of software technology, tools, and platforms that are increasingly becoming indispensable to all domains. Often referred to as `geeks', they display their own unique language, interactions, culture, worldviews, and  social norms.

Software engineers are particularly interesting to study, as they are both the producers and users of technology. They are not only the architects of the software platforms that enable multiple realities, they too function and interact within combinations of these worlds. For example, using a combination of real and digital workspaces, artefacts, and communication mechanisms such as GitHub, physical Scrum task boards, online tools such as Trello or JIRA, physical and virtual pair programming, using Slack channels for team communication, face-to-face and Zoom meetings. All these involve intensively socio-technical interactions.

More generally, depending on the domain and the phenomenon, the actors being studied may be \textit{software engineers} including developers, business analysts, testers, senior managers, or \textit{software users}, including classroom teachers, spacecraft designers, satellite developers, government officials, doctors, nurses, psychologists, artists. Additionally, STGT studies may involve actors that are human, virtual avatars, or AI agents, depending on the nature of the realities where phenomena occur and are studied.

\subsubsection*{Socio-Technical Researcher}

A \textbf{socio-technical researcher} has \textit{the requisite knowledge and skills from social sciences, philosophy, qualitative research, and theory development as well as technical background and/or experience applicable to the domain}. Lack of sociological background is seen as hampering grounded theories \citep{martin2019GT, gibson2013rediscovering}. This does not imply that an ST researcher needs to be an expert in each of these areas. Neither do they have to be one person. To conduct quality STGT studies, a socio-technical researcher, or an interdisciplinary team of researchers, should typically possess a \textbf{core repertoire of knowledge and skills}.

\begin{itemize}
    \item \textit{philosophical foundations} -- understanding fundamental philosophical concepts such as types of reasoning, ontology, epistemology, and research paradigm. A basic overview of these concepts is presented in section \ref{philosophy}.
    \item \textit{qualitative research} -- designing research protocols, ethics assessment/approval, collecting and analysing qualitative data.
    \item \textit{theoretical sensitivity} -- affinity for theoretical abstraction, conceptualisation, and theory development.
    \item \textit{socio-technical sensitivity} -- an affinity for exploring and understanding the intricacies of human interactions and socio-technical aspects of phenomena.
    \item \textit{domain knowledge} -- an understanding of the research domain, typically provided by a subject matter expert.
\end{itemize}

When equipped with relevant background in social sciences and theory development, SE researchers become socio-technical researchers. Those successful in conducting quality GT studies in SE so far (Table \ref{tab:ExemplarGTstudies}) have applied their repertoire of knowledge and skills gained from social sciences, philosophy, theory development, and SE. Researchers lacking the requisite knowledge and skills struggle to achieve quality research outcomes \citep{storey2020EMSE}.

Similarly, sociologists with required technical background are well placed to be socio-technical researchers, conducting STGT studies of phenomenon with prominent and interwoven role of technology. An inter-disciplinary team of software engineering researchers or technologists, sociologists, and domain experts (e.g. banking and finance researchers for studies in the banking domain) can also make for an effective cross-functional socio-technical research team to study an increasingly socio-technical world. The importance of inter-disciplinary and trans-disciplinary research teams is not only relevant to STGT studies, but is also being acknowledged in empirical software engineering in general \citep{fernandez2019empirical}.

\subsubsection*{Socio-Technical Data Tools and Techniques}
Socio-technical researchers studying socio-technical phenomena, domains, and actors can leverage from a range of \textit{traditional and modern research data, tools, and techniques}. 

Traditional data types, sources and collection techniques such as interviews, observations, and field trips offer rich data and insights. However, since the introduction of GT about half a century ago, a number of modern research data, tools, and techniques have emerged. While traditional GT guidelines are not averse or impervious to them, they understandably do not provide explicit guidance or examples of how best to harness them.

SE researchers are at the forefront of the technological revolution in research. Modern research data, tools, and techniques have given rise to entirely new research communities, e.g. mining software repositories. Modern types and sources of data such as publicly available software code repositories with commit messages, source code, documentation, and wikis regularly used in SE research, make for potent new avenues for theory development. Especially, when combined with modern techniques such as data mining and natural language processing, they can be used to extract large amounts of data, hitherto impossible with manual techniques. However, easy access to publicly available datasets and powerful search tools can be tempting to practice `\textit{data-driven research}', where easy access to large datasets precedes and motivates, often trivial, research questions and studies. On the contrary, the use of publicly available, large datasets in STGT is envisioned as a mechanism to scale theories to cover wider contexts of use, while still being firmly grounded in empirical evidence and following systematic and rigorous STGT steps and procedures to iteratively and incrementally derive key patterns from underlying data.

Additional sources of data such as research literature and grey literature are also being explored for theory development \citep{martin2019GT}. In fact, \cite{wolfswinkel2013using} have presented a GT-based approach to conducting literature reviews, called the grounded theory literature review (GTLR). Future STGT studies in SE can explore these avenues to create modern renditions of grounded theories.

Furthermore, traditional GT analysis techniques, \textit{open coding and constant comparison}, can be complemented with modern analysis techniques such as \textit{sentiment analysis} \citep{calefato2018sentiment}, e.g. to study emotions in software engineering. Currently, such innovations are limited to individual studies \citep{madampe2020towards}. In order to influence improvements in GT practice in SE research at scale, the use of modern data, tools, and techniques need to be acknowledged and documented as method guidelines.

Future STGT studies can explore the use of cutting edge tools and techniques such as artificial, virtual, augmented, and mixed reality devices and platforms to achieve \textit{immersive and experiential} data collection in natural and alternative realities. The improving efficacy of automation tools promises to ease and augment many research steps such as transcriptions, data collection, and even analysis. 

\begin{table*}[t]
    \caption{\small Philosophical foundations requisite for socio-technical grounded theory (STGT)}
    \scriptsize
    \centering
    \begin{tabular}{p{1.3cm}p{5.9cm}p{5.9cm}p{3.3cm}}
    \hline\noalign{\smallskip}
    \multicolumn{4}{c}{\cellcolor[gray]{0.9}\textbf{TYPES OF REASONING (AKA INFERENCE OR LOGIC)}}\\
    \hline\noalign{\smallskip}
    \textbf{Reasoning} & \textbf{Description} & \textbf{Example} & \textbf{Manifestation in STGT} \\
    \noalign{\smallskip}
    \textit{Induction}  & A \textit{bottom-up} approach of drawing general conclusions from patterns evidenced across specific instances. Strength of evidence improves certainty of the conclusion that is \textit{localised not universal}. &  A sparrow in a city is observed with brown-black-white feathers, another sparrow is observed with similar coloured feathers, and another, until the observer finds a pattern amongst a number of sparrows and concludes that all sparrows in the city are likely to have brown-black-white feathers. & Open coding, targeted coding, and constant comparison. \\
    \noalign{\smallskip}
    \textit{Deduction}  & A \textit{top-down} approach of drawing conclusions about specific instances from general facts or theories. The conclusion can be verified through repeated testing, giving rise to the term \textit{hypothetico-deductive} approach, and is guaranteed certainty until proven false. & All birds have beaks, a Pigeon is a bird, so a Pigeon has a beak. & Theoretical sampling. \\
    \noalign{\smallskip}
    \textit{Abduction}  & A `\textit{detour}' approach of proposing the best possible explanation based on all the evidence, albeit incomplete, available. Abduction plays an important role in theory development. & If bird feathers are spotted on the doormat and the family cat is seen with some feathers on their paws, the most likely explanation is that the cat ate the bird. & Memoing, theoretical structuring, theoretical integration.\\
    \end{tabular}
    \label{tab:typesofreasoning}
\end{table*}

\begin{table*}[]
    \centering
    \scriptsize
    \begin{tabular}{p{1.1cm}p{3.9cm}p{6.4cm}p{5cm}}
    \hline\noalign{\smallskip}
    \multicolumn{4}{c}{\cellcolor[gray]{0.9}\textbf{ONTOLOGY, EPISTEMOLOGY, \& RESEARCH PARADIGMS}}\\
    \hline\noalign{\smallskip}
    \textbf{Concept} & \textbf{Definition} & \textbf{Type \& Example} & \textbf{Manifestation in STGT}\\
    \noalign{\smallskip}
    \textit{Ontology} & Ontology refers to what we believe exists or what we perceive as reality, in a research context. & Only directly observable material bodies are real, or both material and immaterial forms are real, or reality includes socially constructed concepts such as identities and beliefs. & STGT enables the study of physical, virtual, combined, and simulated realities.\\
    \noalign{\smallskip}
    \textit{Epistemology} & Epistemology refers to what we treat as knowledge and how we can know about what we perceive as reality, in a research context. & A \textit{subjectivist} epistemology holds that knowledge about actors \& phenomena is subjectively constructed through inputs from the participants by non-interchangeable researchers. An \textit{objectivist} epistemology holds that knowledge about objects \& phenomena is derived through neutral measurements or observations by interchangeable researchers. & Subjective epistemology for physical, virtual, and combined contexts. Objective epistemology possible for some simulated contexts.\\
    \noalign{\smallskip}
    \textit{Research Paradigm} & Research paradigm refers to the combination of ontological and epistemological stances that informs the researcher's worldview. & \textit{Constructivism} aligns with the concept of a socially constructed reality that is subjectively co-constructed. \textit{Positivism} aligns with the concept of a standalone external reality that can be objectively observed. & Constructivism with Symbolic interactionism in physical, virtual, \& combined contexts. Positivism in simulated contexts. Combinations possible (e.g. constructivist-feminist GT)\\

    \hline\noalign{\smallskip}
    \end{tabular}
\end{table*}

\subsection{STGT -- Philosophical Foundations}
\label{philosophy}
This section presents the fundamental philosophical concepts requisite to conducting quality STGT studies including types of reasoning, ontology, epistemology, and research paradigms. These are based on Peircean \citep{peirce1960collected, peircePlato, abductionPlato}, more modern philosophical views \citep{guba1994competing} and relevant software engineering literature \citep{russo2001use, peggy2008selecting, wohlin2015towards, seaman1999qualitative} and are presented in an accessible way for researchers non-native to social science. Table \ref{tab:typesofreasoning} presents an overview of these key concepts including simple definitions, examples, and how they are manifested in research. Table \ref{tab:comparison} presents a comparison of STGT with traditional Glaserian, Strauss-Corbinian, and Constructivist GT, across their philosophical foundations (second block). Understanding these fundamental philosophical concepts is critical to performing quality research studies and evaluations, especially STGT studies.

\subsubsection*{Types of Reasoning}

GT challenged the top-down \textit{hypothetico-deductive} approach to verifying existing theories dominating the sociological research scene in the mid-1960's and steered it toward a bottom-up \textit{inductive} approach to generating new and original theories. In a bid to delineate its stance, traditional GT overemphasises the role of induction but classifying GT as purely inductive is na\"ive \citep{haig1995grounded}. Different types of reasoning, \textit{inductive}, \textit{deductive}, and \textit{abductive} are seen to apply in theory development. Please refer to Table \ref{tab:typesofreasoning} for their simple definitions and examples. 

The role of \textit{deduction} in guiding theoretical sampling is acknowledged in \textit{Glaserian GT} \cite[chapter~II]{glaser2017discovery}. \textit{Strauss-Corbinian GT} suggests a greater role of deduction in the later stages of a GT study as the emergent theory is systematically verified against the data \citep{strauss1990basics} while \textit{Glaserian GT} is strongly opposed to this approach, claiming it as ``\textit{forcing}'' rather than emergence of theory \citep{glaser1992basics}.

On the other hand, the role of sudden or slow-dawning ``\textit{sensitive insights of the observer}'' is acknowledged as ``\textit{the root sources of all significant theorizing}'' \citep[chapter~XI]{glaser2017discovery}. While the descriptions of these ``insights'' are close to the definition of \textit{abduction} by Charles Sanders Peirce who defines it as ``\textit{a process of forming an explanatory hypothesis}'' \citep{abductionPlato, haig1995grounded}, \textit{Glaserian GT} does not explicitly acknowledge the role of abduction in theory development. \textit{Strauss-Corbinian GT} is said to have been influenced by abduction but its reference remains limited to a footnote in Strauss' book \citep{charmaz2014constructing, strauss1987qualitative}.

Charmaz acknowledges abduction as a type of ``inferential leap'' required of the GT researcher when they arrive at a surprising finding that does not fit the emergent patterns \citep{richardson2006abduction, charmaz2014constructing}. Using this ``imaginative way'' of reasoning, the researcher comes up with useful explanations (\textit{theoretical conjectures} or inferences) in an attempt to account for the surprising finding, then returns to re-examine or collect more data to check them \citep{charmaz2014constructing}. 

The role of abduction in developing grounded theories remains to be fully explained \citep{bruscaglioni2016theorizing, martin2019GT}. Understanding different abductive reasoning modes such as \textit{reason or hunch, clue, metaphor or analogy, symptom, pattern,} and \textit{explanation} \citep{shank1998extraordinary} opens a variety of avenues for creative thinking which is core to theorising and theory development \citep{martin2019GT}. For example, abduction is the basis for diagnosis in medical sciences, where plausible hypotheses are explored and then accepted or discarded based on how well they `fit', i.e. explain the symptoms being observed. Criminal investigations also offer examples of the three types of reasoning, particularly abduction in action, most prominently demonstrated in the engaging pursuits the fictional character of Sherlock Holmes, but also referred to by Peirce from a personal `detective' experience \citep{reichertz2007abduction}. Similarly, abduction is applied in research for theory development to suggest hypotheses based on evidence. Through iterative data collection, analysis, and \textit{memoing}, these hypotheses fall through or become established over the course of the study. Additionally, coming up with robust hypotheses about a socio-technical phenomenon requires a full understanding of the socio-technical context.

STGT, like traditional GT, follows a predominantly and overarching inductive approach, drawing generalised conclusions from patterns evidenced across specific instances. Induction in STGT is most clearly manifested in its \textit{open} and \textit{targeted coding} and \textit{constant comparison} procedures where evidence is repeatedly raised in levels of abstraction toward a theory. STGT acknowledges and explains the role of deduction as applied in \textit{theoretical sampling} and, unlike traditional GT (see Table \ref{tab:comparison}), the role of abduction as applied in \textit{memoing}, \textit{theoretical structuring}, and \textit{theoretical integration} (procedures described later.)

\subsubsection*{Ontology}
Ontology refers to \textit{what we believe exists or what we perceive as reality, in a research context}. As researchers, we hold different views about what constitutes reality in the research context, referred to as our ontological stance. For example, we may believe only directly observable material bodies in the natural world constitute reality that is certain or that reality includes socially constructed concepts such as identity, culture, and beliefs, and  combinations thereof. Researchers inclined to study the former (material reality) are often seen to prefer what are typically referred to as `hard' sciences such as physics and maths, where the reality of the research context includes relatively stable phenomena (e.g. acceleration, gravity) studied through observing or measuring actions on interchangeable objects (e.g. apples), typically using quantitative research (numbers, accuracy, precision). Researchers inclined to study the latter (socially constructed concepts as reality) are often seen to prefer what are traditionally referred to as `soft' sciences such as psychology and sociology, where the reality of the research context includes phenomena that are full of variations (e.g. driving, cooking, programming) studied through interactions with non-interchangeable subjects (e.g. people), typically using qualitative research.

Software engineering research predominantly involves studying socio-technical phenomena that unfold in multiple realities, beyond the traditional physical reality, e.g. in artificial, virtual, augmented, and combinations of realities. For example, studying a typical software development team involves physical contexts (e.g. physical scrum boards) and physical interactions (e.g. physical daily standups) as well as virtual contexts (e.g. project management on JIRA or Trello), virtual interactions (e.g. video calls, emails, virtual chat, online discussion forums), and even combined interactions (e.g. two developers collaborating in-person over a virtual task board self-assigning tasks using symbols such as their online avatars). Contexts such as remote work are primarily virtual, offering different affordances, norms, and symbolism (e.g. ``\textit{you are on mute}'' makes no sense in a physical meeting). Studying social interactions between real people using their online profiles and avatars on online social media platforms such as Twitter and Facebook presents yet another type of reality as real versus fake identities come into play (e.g. symbols such as the `blue tick' on Twitter enabling authentication.) These multiple realities are defined by intermixing operational contexts, symbols, languages, norms, and guidelines of both physical and virtual worlds, offering same if not more complexity than the physical world. 
While traditional GT guidelines address physical realities, STGT acknowledges the diverse and combined ontological frames prevalent in SE and provides guidelines to enable the study of physical, virtual, combined, and simulated realities.

\subsubsection*{Epistemology}
Epistemology refers to \textit{what can be treated as knowledge and how that knowledge is gained, in a research context.} As researchers, we hold different views about how we can gain knowledge about the reality of our research context, referred to as our epistemological stance. For example, we may believe that we can learn about the research world (objects and phenomena therein) through neutral measurements or observations where exchanging an observer with another will yield the same results. This is referred to as an \textit{objectivist} epistemology. On the other hand, we may believe that we can learn about the research world through subjective interpretations of observations where exchanging the observer with another will likely yield different findings. This is referred to as a \textit{subjectivist} epistemology.

STGT highlights that a researcher's epistemological stance is typically determined by two aspects, the nature of the reality being studied, e.g. physical,  virtual, augmented, combined, and the researcher's own preferences. For example, STGT studies of virtual game worlds can be designed with an \textit{objectivist} approach where the game analytics are used to collect objective metrics about the game (e.g. lives lost, levels successfully cleared, time taken per level, choice of actions and interactions). Similarly, STGT studies of combined realities, such as development team environments, can apply a mix of objective metrics and observations that are interpreted subjectively.

\subsubsection*{Research Paradigm}

A combination of ontological and epistemological stances form a research worldview. Research paradigms \textit{formally capture the researcher worldview about what they believe is reality (\textit{ontology}) and how knowledge about that reality can be gained (\textit{epistemology}), in a research context}.  Popular paradigms include positivism, post-positivism, constructivism, symbolic interactionism, and pragmatism. Let us consider the prominent paradigms associated with GT below, compared under \textit{philosophical foundations} in Table \ref{tab:comparison}.\\

\noindent\textbullet\space\space \textit{\textbf{Positivism}} refers to \textit{a definite, assured, and certain nature of reality that can be studied through direct observations, and excludes traditional metaphysics and theology}. The ontological lens used in positivism suggests reality is external and standalone, irrespective of human existence or interference, and ready to be discovered through (only) what can be observed and measured. The observer's job is to discover and render it accurately. The epistemology native to positivism is \textit{objectivist} which suggests knowledge about reality is discovered through objective observations and measurements by neutral and interchangeable observers who do not influence the reality by their presence or measurements. Consequently, positivists emphasise verification through testing of hypotheses or existing theories against observable facts and value reproducibility, replicability, refutability, generalisability, and statistical significance. 
Research methods typically following a positivist paradigm include controlled and quasi-experiments, although survey and case study research is also known to be conducted with a positivist approach \citep{peggy2008selecting}.
    
 Positivism is naturally associated with \textit{quantitative data} and \textit{deductive reasoning}, drawing specific conclusions from general facts through repeated testing and proving/falsification of hypotheses. Historically, quantitative researchers with positivist inclinations have been staunch critics of qualitative research often conducted with interpretive approaches \citep{charmaz2014constructing}. \textbf{Post positivism}, on the other hand, refocuses attention on falsification of hypothesis, increasing confidence in proposed theories with every failed attempt at falsification \citep{peggy2008selecting}.

Classic or Glaserian GT has been associated with a positivist approach \citep{charmaz2006constructing}.\\

\noindent\textbullet\space\space \textit{\textbf{Constructivism}} \textit{supports the ontology of a reality that is socially constructed as (opposed to being external, standalone reality) and an epistemology that supports the view that the researchers' and participants' presence and interactions construct the reality that is studied (as opposed to some objective and neutral discovery of an external reality)}. In other words, ``\textit{knowledge and truth are created, not discovered by mind}'' \citep{schwandt1994constructivist}. Because of its native \textit{subjective} stance, researchers are not considered neutral or interachangeable, rather mediums through which observations are made and interpreted in unique ways. Consequently, findings from a research study conducted using an constructivist approach are acknowledged to be limited to the same or similar contexts in which they were conducted and not universally generalisable.

Constructivists emphasise generation of new knowledge or theories over verifying existing knowledge or theories, and value originality, novelty, usefulness, and real-world significance. Research methods most closely aligned with a constructivist approach include ethnography and case studies \citep{peggy2008selecting}.  

The primary premise of introducing a new version of GT by Charmaz was to introduce and apply a constructivist research approach to conducting GT \citep{charmaz2006constructing}.\\

\begin{table*}[]
    \scriptsize
    \caption{\small Socio-Technical Grounded Theory (STGT) compared to traditional Glaserian GT (1967), Strauss-Corbinian GT (1990), and Constructivist GT (2006) methods across research context, philosophical foundations, methodological \& evaluation guidelines.}
    \label{tab:comparison}
    \begin{tabular}{>{\raggedright\arraybackslash}p{2.7cm}>{\raggedright\arraybackslash}p{2.4cm}>{\raggedright\arraybackslash}p{4.2cm}>{\raggedright\arraybackslash}p{3.6cm}>{\raggedright\arraybackslash}p{3.3cm}}
    \hline\noalign{\smallskip}
    \multicolumn{5}{c}{\cellcolor[gray]{0.85}\textbf{RESEARCH CONTEXT}}\\
    \hline\noalign{\smallskip}

    \textbf{Research Method} & \textbf{Phenomena} & \textbf{Domain} & \textbf{Data, Tools \& Techniques} & \textbf{Researcher}\\
    \noalign{\smallskip}\hline\noalign{\smallskip}
    {\cellcolor[gray]{0.94}\textit{\textbf{Socio-technical GT}} (Hoda, 2020)}  & {\cellcolor[gray]{0.94}\textit{Socio-technical}, with interwoven social and technical aspects (e.g. ethics in AI).} & {\cellcolor[gray]{0.94}\textit{Socio-technical}, software engineering (original), computer science, artificial intelligence, human computer interaction, information systems, and other domains with varying levels of social and technical interplay.} & {\cellcolor[gray]{0.94}Qual data (primary), Quant data (supplementary). \textit{Traditional} e.g. interviews, observations, and \textit{Modern}, e.g. data mining, NLP, sentiment analysis, immersive techniques.} & {\cellcolor[gray]{0.94}\textit{Socio-technical researcher/team}, with technical/domain knowledge \& relevant skills in social sciences, philosophy, qualitative research, \& theory development.}
    \\
    \noalign{\smallskip}
    \textit{Traditional GT methods} (Classic, Strauss-Corbinian, Constructivist) & \textit{Social}, primary focus on  humans and social interactions (e.g. social loss). & \textit{Various}, medicine (original), nursing, psychology, education, attempted in software engineering with varying success. & Qual data (primary), Quant data (supplementary). \textit{Traditional}, e.g. interviews and observations. & \textit{Sociologist/team}, with knowledge and skills in social sciences, philosophy, qualitative research, and theory development.\\
    \end{tabular}

    \label{tab:misalignment}
\end{table*}

\begin{table*}[]
    \centering
    \scriptsize
    \begin{tabular}{>{\raggedright\arraybackslash}p{2.7cm}>{\raggedright\arraybackslash}p{2.4cm}>{\raggedright\arraybackslash}p{2.4cm}>{\raggedright\arraybackslash}p{3.2cm}>{\raggedright\arraybackslash}p{5.4cm}}
    \hline\noalign{\smallskip}
    \multicolumn{5}{c}{\cellcolor[gray]{0.85}\textbf{PHILOSOPHICAL FOUNDATIONS}}\\
    \hline\noalign{\smallskip}
    \textbf{Research Method}  & \textbf{Reasoning} &  \textbf{Ontology} & \textbf{Epistemology} & \textbf{Research Paradigm}\\
    \noalign{\smallskip}\hline\noalign{\smallskip}
    {\cellcolor[gray]{0.94}\textit{\textbf{Socio-Technical GT}} (Hoda, 2020)} &  {\cellcolor[gray]{0.94}Inductive, deductive \& abductive}  & {\cellcolor[gray]{0.94}Context-specific, e.g. physical, virtual, combined, simulated.} & {\cellcolor[gray]{0.94}Context specific, e.g. subjective in natural settings, objective in simulated world.} & {\cellcolor[gray]{0.94}Context specific, e.g. constructivism in natural and virtual worlds, positivism in simulated worlds.} Combinations possible. \tabularnewline
    \noalign{\smallskip}
    \textit{Constructivist GT} (Charmaz, 2006) & Inductive \& abductive & Physical (original) & Subjective & Constructivism \\
    \noalign{\smallskip}   
    \textit{Strauss-Corbinian GT} (Strauss \& Corbin, 1990)  & Inductive \& deductive & Physical (original) & Interpretive & Symbolic interactionism \\
    \noalign{\smallskip}   
    \textit{Glaserian GT} (Glaser \& Strauss, 1967)  & Inductive (\& hints of deductive) & Physical (original) & Objective (original) & Positivism \\
    \end{tabular}
\end{table*}

\begin{table*}[]
    \centering
    \scriptsize
    \begin{tabular}{>{\raggedright\arraybackslash}p{2.7cm}>{\raggedright\arraybackslash}p{3cm}>{\raggedright\arraybackslash}p{4.8cm}>{\raggedright\arraybackslash}p{5.8cm}}
    \hline\noalign{\smallskip}
    \multicolumn{4}{c}{\cellcolor[gray]{0.85}\textbf{METHODOLOGICAL STEPS AND PROCEDURES}}\\
    \hline\noalign{\smallskip}
    \textbf{Research Method}  & \textbf{Literature Review} & \textbf{Data Collection} &  \textbf{Data Analysis \& Theory Development} \\
    \noalign{\smallskip}\hline\noalign{\smallskip}
    
    {\cellcolor[gray]{0.94}\textit{\textbf{Socio-Technical GT}} (Hoda, 2020)} 
    & {\cellcolor[gray]{0.94}\textit{Lean} literature review, \textit{Targeted} literature review.} 
    & {\cellcolor[gray]{0.94}\textit{All is data, with care} -- qual. primary (quan. supplement) from \textit{credible} sources with \textit{ethical} considerations, using traditional} \& {\cellcolor[gray]{0.94}modern approaches via different sampling  initially \& theoretical sampling later.}
    & {\cellcolor[gray]{0.94}\textbf{Basic Stage} -- open coding, constant comparison, basic memoing. \textbf{Advanced Stage} -- advanced memoing and option (1) \textit{Emergent mode} (targeted coding, constant comparison, theoretical structuring) \textit{or} option (2) \textit{Structured mode} (structured coding, constant comparison, theoretical integration).}\\
    \noalign{\smallskip}
    
    \textit{Constructivist GT} (Charmaz, 2006) & Customised literature review. & Variety of elicited and extant qual. data via initial \& theoretical sampling. & Initial coding, focused coding, theoretical coding. \\
    
    \noalign{\smallskip}   
    \textit{Strauss-Corbinian GT} (Strauss \& Corbin, 1990) & Multiple \& extensive use of literature review. & Variety of primarily qual. data via theoretical (open, relational, discriminate) sampling. & Open coding, constant comparison, axial \& selective coding, memoing, sorting. \\
    
    \noalign{\smallskip}   
    \textit{Glaserian GT} (Glaser \& Strauss, 1967) & Avoid literature reviews in same area.  & \textit{All is data} -- qual. primary (quant. possible) using interviews \& observations via theoretical sampling. & Open coding, constant comparison, selective coding, memoing, theoretical sorting, theoretical coding.  \\
    \end{tabular}
\end{table*}

\begin{table*}[]
    \centering
    \scriptsize
    \begin{tabular}{>{\raggedright\arraybackslash}p{2.7cm}>{\raggedright\arraybackslash}p{4.8cm}>{\raggedright\arraybackslash}p{4cm}>{\raggedright\arraybackslash}p{5cm}}    
    \hline\noalign{\smallskip}
    \multicolumn{4}{c}{\cellcolor[gray]{0.85}\textbf{EVALUATION GUIDELINES}}\\
    \hline\noalign{\smallskip}
    \textbf{Research Method}  & \textbf{Evaluating Method Application} &  \textbf{Evaluating Partial Findings} &  \textbf{Evaluating Mature Theories}\\
    \noalign{\smallskip}\hline\noalign{\smallskip}
    {\cellcolor[gray]{0.94}\textit{\textbf{Socio-Technical GT}} (Hoda, 2020)} & {\cellcolor[gray]{0.94}Credibility, rigour} & {\cellcolor[gray]{0.94}Originality, relevance, density} & {\cellcolor[gray]{0.94}Novelty, usefulness, parsimony, modifiability}  \tabularnewline
    \noalign{\smallskip}
    \textit{Constructivist GT} (Charmaz, 2006) &  - & - & Credibility, originality, resonance, usefulness \\
    \noalign{\smallskip}   
    \textit{Strauss-Corbinian GT} (Strauss \& Corbin, 1990) & Sampling, major categories, indicators, theoretical sampling, hypotheses, discrepancies, core category & - & Concepts, categories, and links generation, variation, process reliability, significance \\
    \noalign{\smallskip}   
    \textit{Glaserian GT} (Glaser \& Strauss, 1967)  &  - & - & Fit, work, relevance, modifiability   \\
    \hline\noalign{\smallskip}
    \end{tabular}
\end{table*}

\noindent\textbullet\space\space \textit{\textbf{Other paradigms, perspectives, theories}} include \textbf{pragmatism} \citep{james1975pragmatism}, which \textit{aligns with the ontology that sees reality as what is of practical use and an epistemology that is based on understanding reality in ways that address what practical difference it might make to the observer}. \textbf{Symbolic interactionism}, often referred to as a theory rather than a paradigm \citep{schwandt1994constructivist}, is guided by the ontological view that reality is what is socially understood and the epistemological view that reality is perceived through social interactions that define and redefine symbolic meaning of concepts. Strauss-Corbinian GT is characterised as building on pragmatism and symbolic interactionism underpinnings \citep{strauss1990basics}.

In addition to research paradigms, some unique perspectives can be applied to research. \textbf{Feminism} is concerned with the representation of females in the studied phenomena and so applies that lens to all aspects of research including research design, data collection, analysis, presentation, and evaluation. \textbf{Critical theory} is based on the belief that research serves a political purpose and can be used to empower specific groups, specially minority groups through raising awareness and recommending change.

While no one research paradigm or worldview is correct or wrong, it is important for researchers to be aware of and declare their philosophical stance because it influences how we conduct studies, present findings, and most prominently, how we evaluate research.  For example, a positivist reviewer evaluating a research study conducted using a constructivist approach is likely to -- consciously or unconsciously -- apply positivist evaluation criteria, inappropriate for the study, e.g. look for generalisability instead of novelty or expect reproducibility and replicability instead of credibility and rigour, and vice-versa. Further information on research paradigms and their impact on research design can be consulted in \citep{seaman1999qualitative, peggy2008selecting, wohlin2015towards, russo2001use}. The use of research paradigms in SE research has been explored in \citep{melegati2021surfacing}.

 An interesting point to note is that an individual's approach to life in general, i.e. their general worldview, while likely to influence their research worldview, need not be the same. For example, an individual with a pragmatic approach to making their life choices may consciously chose to apply a positivist approach to their research study.

 More often, researchers prefer a research paradigm over another and resort to applying their preferred paradigm for a majority of their research studies. However, researchers need not be bound to any one research paradigm for all their studies and can chose to conduct different studies using paradigms best suited to the nature of the study. Such flexibility requires a mature understanding of philosophical foundations and high levels of researcher self-awareness and reflection. Similarly, research paradigms and perspectives can be combined, e.g. critical positivist or feminist-constructivist to achieve unique research aims and flavours.

\subsubsection*{STGT Research Paradigms}

Traditional GT methods have been associated with particular paradigms, Glaserian GT with positivism \citep{charmaz2006constructing}, Strauss-Corbinian as building on pragmatism and symbolic interactionism \citep{strauss1990basics}, and Charmaz's GT with constructivism \citep{charmaz2006constructing}.

STGT, as the name suggests, applies a socio-technical worldview to conducting GT research, which is open to selecting and applying specific ontological and epistemological stances as best suited to the research context. In other words, STGT can be applied using different research worldviews based on the ontology applicable to the study and on the researcher’s preference. This does not equate to being paradigm ignorant or agnostic, rather it requires a careful consideration of the research context to ascertain the choice of paradigm to apply.

 For novice researchers, some suggestions are provided. For example, STGT studies of physical and/or combined realities, such as offered by software team environments, may be conducted using a subjective approach where the nature of the findings are dependent on the researcher conducting the data collection and analysis. On the other hand, STGT studies of simulated and virtual realities, such as in computer games, may apply an objective approach, leveraging in-game analytics. Studying end-user interaction with AI manifested as chatbots or robots offers similar contexts. Such simulated and combined realities can be designed to be finite and deterministic, manifesting closely the idea of an external standalone reality with limited possible interpretations, best captured by a objectivist stance.

 Some research areas can be studied using different worldviews depending on the study focus and objectives. For example, a study on test case selection can be conducted using an objective approach if focusing purely on the technical tools and their performance, validating and verifying outcomes against goals. While a constructivist approach can be applied to the same research context if considering the tester's motivations, rationales, and behaviours with regards to test case selection. STGT studies can also combine a selected epistemology with established theories and perspectives used as an additional lens, e.g. a constructivist-feminist STGT study of female contributors to open source software, their experiences, challenges, and strategies.

By acknowledging the research paradigms and perspectives,  researchers can remain aware and consistent in their application, and actively set the expectations for their readers and reviewers so that the findings can be understood and evaluated in commensurate ways. 

\begin{figure*}[t]
    \centering
    \includegraphics[width=1\textwidth]{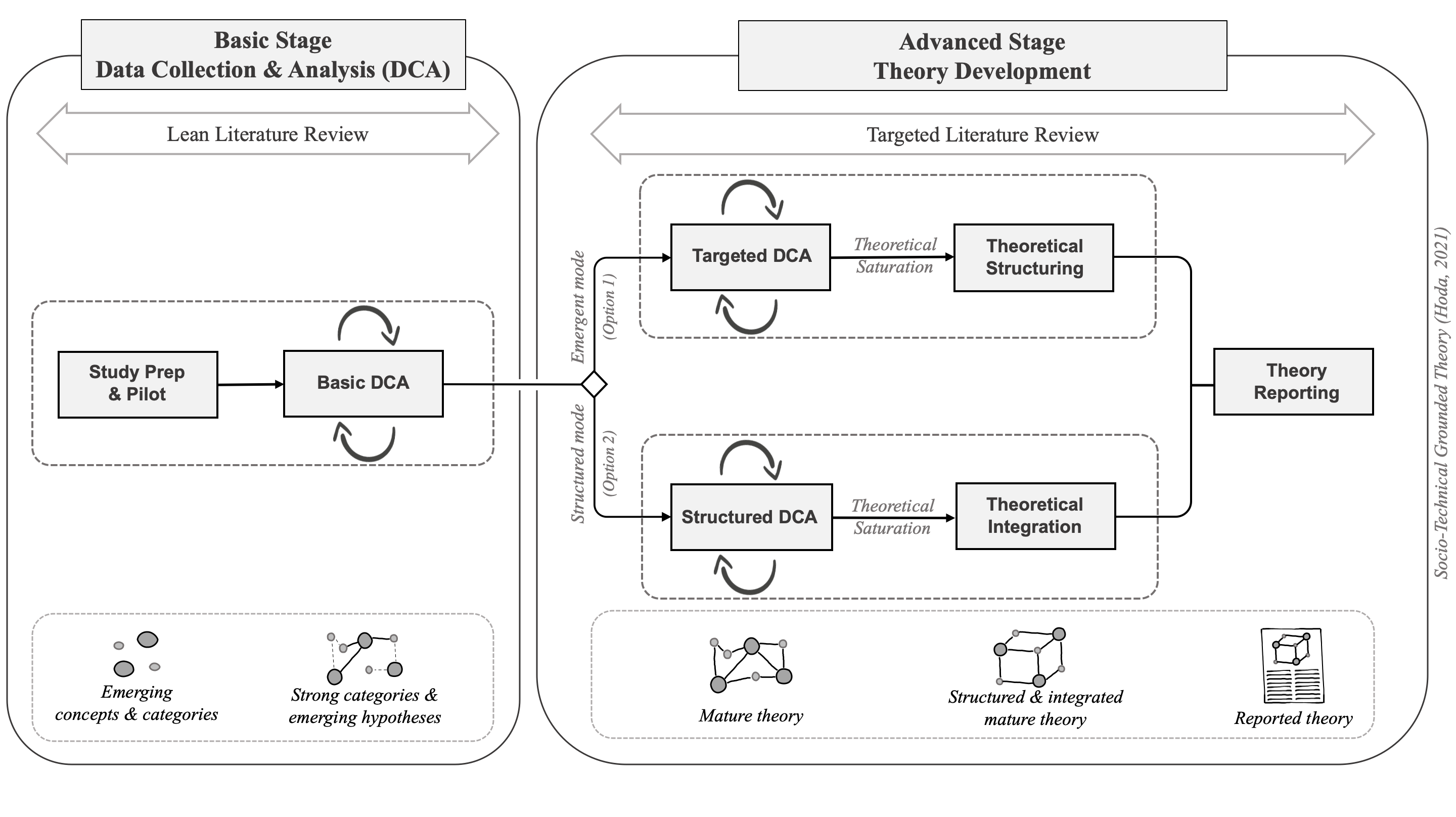}
    \vspace{-0.8cm}
    \caption{Socio-Technical Grounded Theory Method -- key stages (two large blocks of basic and advanced stages, labelled at the top), steps and procedures (middle) including options of emergent and structured modes of theory development, and the progressively emerging outcomes (bottom row)}.
    \label{fig:STGT}
\end{figure*}

\section{STGT Method -- Overview and Basic Stage}
\label{method}
These next sections present an overview of the Socio-Technical Grounded Theory method and describe its methodological steps and procedures. Further details of the STGT method introduced in this article can be found in the author's forthcoming book. Example applications of specific steps and procedures (e.g. open coding, memoing) are referenced throughout.

\subsection{STGT -- An Overview}
 The STGT method comprises of steps and procedures adapted from the three different versions of traditional GT -- \textit{Glaserian}, \textit{Strauss-Corbinian}, and \textit{Constructivist} -- in a way that eases entry into an STGT study through a \textit{\textbf{basic data collection and analysis stage}} and offers flexibility through two options in the \textit{\textbf{advanced theory development stage}}, following either an emergent approach to theory development, similar to \textit{Glaserian GT}, or a structured approach, similar to \textit{Strauss-Corbinian GT}. It also offers flexibility in the choice of ontology, epistemology, and therefore, research paradigm applied as best suited to the study context, including the constructivist paradigm as applied in \textit{Constructivist GT}.

 Table \ref{tab:comparison} provides a comparison of STGT with the traditional GT methods, including similarities and adaptations. For example, STGT's approach to literature review is closer to the \textit{Constructivist GT} approach, looking for a balance between being sufficiently informed versus overly influenced by existing works. STGT's approach to progressively narrow down the scope of the data collection and analysis to focus on the key emerging categories over time reflects the same fundamental trajectory of all traditional GT methods as they move from open to some form of targeted data collection and analysis. Like \textit{Strauss-Corbinian GT}, STGT distinguishes between evaluating the research method application and outcomes. Additionally, STGT also acknowledges partial findings and lists criteria to evaluate them differently to mature findings.

Figure~\ref{fig:STGT} presents a visual model or diagram of the socio-technical grounded theory method including its key stages, depicted within two large blocks of \textit{basic} and \textit{advanced} stages (labelled at the top), steps and procedures (middle row) including options of \textit{emergent} or \textit{structured} modes of theory development, and outcomes progressively emerging from the steps (bottom row).\\

\noindent \textbf{Basic stage -- Data Collection and Analysis (DCA)}, involving the steps of \textit{lean literature review, study preparation and piloting}, and iterations of \textit{basic data collection and analysis}. Initial concepts and categories start to emerge from the piloting (see bottom row of the model diagram). The development of a few strong categories and preliminary hypotheses mark the end of the basic stage. \\
    
\noindent \textbf{Advanced stage -- Theory Development}, involving \textit{targeted literature review} as and when needed, and selecting from two modes of theory development:
\begin{itemize}
    \item \textbf{Emergent mode}, enabling the emergence of theory through the iterative \textit{targeted data collection and analysis} step which ends with \textit{theoretical saturation} and resulting in a mature theory that is integrated. The theory can be further structured using theoretical presentation templates during \textit{theoretical structuring}.
    \item \textbf{Structured mode}, enabling a structured development of theory through \textit{structured data collection and analysis} ending with  \textit{theoretical saturation} and resulting in a mature theory that is structured and can be further integrated through \textit{theoretical integration}.
\end{itemize}

The term \textit{development} here is used for theory generation or creation. Emerging categories and hypotheses can be reported as interim and mature theory as final findings.

\subsection{Literature Review} 

STGT offers two types of literature reviews, \textit{lean literature review} and \textit{targeted literature review}, at the basic and advanced stages of the research project respectively. 

\begin{itemize}
\item \textbf{Lean literature review (LLR)}, a lightweight and high-level review performed early in the study, during the basic stage, to identify research gaps and motivate the need for a study. An LLR may identify that the topic is relatively nascent, with no or few existing theories, and can benefit from an original theory in the area. Alternatively, it may identify that the topic is extensively studied but lacks theories, or the existing theories do not resonate with practical experience (from reading practitioner literature or personal experience as a practitioner in the field). A GT study in the latter case should be attempted by experienced theorists. Where an LLR identifies robust existing theories on the topic, changing or refining the topic is suggested.\\

\item \textbf{Targeted literature review (TLR)}, an in-depth review of literature targeting relevance to the emerging/emergent categories and hypotheses, performed periodically during the advanced stage. A TLR helps compare the emergent original findings with existing work and situate them in the wider research landscape, filling research gaps.
\end{itemize}

\noindent STGT offers three unique advantages of literature reviews.

\textbf{Improving theoretical sensitivity.} Learning about theories and theory development outside of the study domain is highly recommended to \textit{improve theoretical sensitivity}. For example, SE researchers who do not possess relevant training in sociology or theorising will benefit from exploring theories and theory development in other fields such as sociology, psychology, and nursing.

\textbf{Improving socio-technical sensitivity and domain knowledge.} Reading practitioner and research literature within the study domain will improve \textit{domain knowledge} and \textit{socio-technical sensitivity} -- the ability of the researcher, e.g. non-native or relatively new to the domain, to understand its terminology, concepts, and associated jargon. Domain knowledge and socio-technical sensitivity enables useful and effective data collection, analysis, and reporting.

\textbf{Refining or advancing the state of theory.} This is an advanced practice. With experience in the study domain and in theorising, researchers are likely to become aware of the relevant theories in their field. This knowledge can motivate the need for a fresh look at familiar problems for which a number of solutions, models, or theories may exist, in an attempt to \textit{refine or advance the state of theory}. At the same time, they will need to ensure that thorough knowledge or reviews of relevant literature does not adversely influence their ability to develop original theories. The \textit{theory of becoming agile} \citep{hoda2017becoming} is an example of this principle in action, advancing the state of theory on agile transformations in software teams.\\

\noindent \textbf{Systematic Reviews}. Formal and comprehensive literature reviews popular in SE, such as \textit{systematic literature reviews} (SLR) and \textit{mapping studies} \citep{kitchenham2004procedures}, are \textit{not} recommended for \textit{novice} researchers, those new to research in general or those new to GT. However, these are not overruled for experienced researchers. Systematic reviews and mapping studies primarily apply a positivist paradigm, aiming for completeness, replication of the process, and reproducibility of the results. Where a  systematic review is performed in/with an STGT study, the onus is on the researcher(s) to be aware of, manage, and explain the interplay between an SLR's inherent positivist stance and the stance adopted for the STGT study. They also need to explain the  relationship between the findings of the SLR and the subsequent original STGT findings. For example, if applying an emergent mode of theory development, it is important that the researcher(s) explain the role of the systematic review with regards to (a) the mechanisms used to avoid being biased by the review findings, if using a positivist approach or (b) how it informed the construction of the theory, if using a constructivist approach. On the other hand, the themes or taxonomies derived from the systematic review can be used to drive a structured mode of theory development.\\

\noindent  \textbf{Grounded Theory Literature Reviews}. As mentioned earlier under additional sources of data in section \ref{context}, a GT approach to literature review is also possible. Future STGT studies in SE can attempt the \textit{Grounded Theory Literature Review} method as proposed by \cite{wolfswinkel2013using}. Due to its methodological alignment, a GTLR may be better suited to accompany an STGT study as compared to an SLR.

\subsection{Study Preparation and Piloting}
In preparation for an STGT study, as with most other studies, the researcher(s) needs to select a research topic, a research team, prepare the study protocols, conduct pilot data collection and analysis, and apply necessary refinements. The research team will benefit from including an experienced theorist as a core member or advisor but can also be comprised of members new to theory development, following these STGT guidelines. 

Depending on the data collection techniques envisioned, an initial set of research protocols in the form of recruitment strategies, semi-structured interview questions, online surveys or questionnaires, observation protocols, and modern data mining techniques can be designed. An ethical assessment should be conducted, and if applicable, approved by an ethics committee. Once approved, a handful of pilot data collection and analysis instances can be carried out. For example, new researchers will benefit from one or two pilots to test the study protocols followed by necessary refinements, and two or three others to enable familiarity with data collection and analysis procedures. Experienced researchers will also gain from piloting as each topic presents new contexts and challenges.

Data collected as part of piloting can be used in the study unless the piloting was done in contexts different to the actual study (e.g. piloting a survey with research students where the study focuses on industry practitioners) or the protocols needed to be completely redesigned, including changing the research topic, as can happen in rare instances.\\

\subsection{Basic Data Collection and Analysis}
\begin{wrapfigure}[12]{l}{0.25\textwidth}
    \centering
    \vspace{-10pt}
    \includegraphics[width=0.25\textwidth]{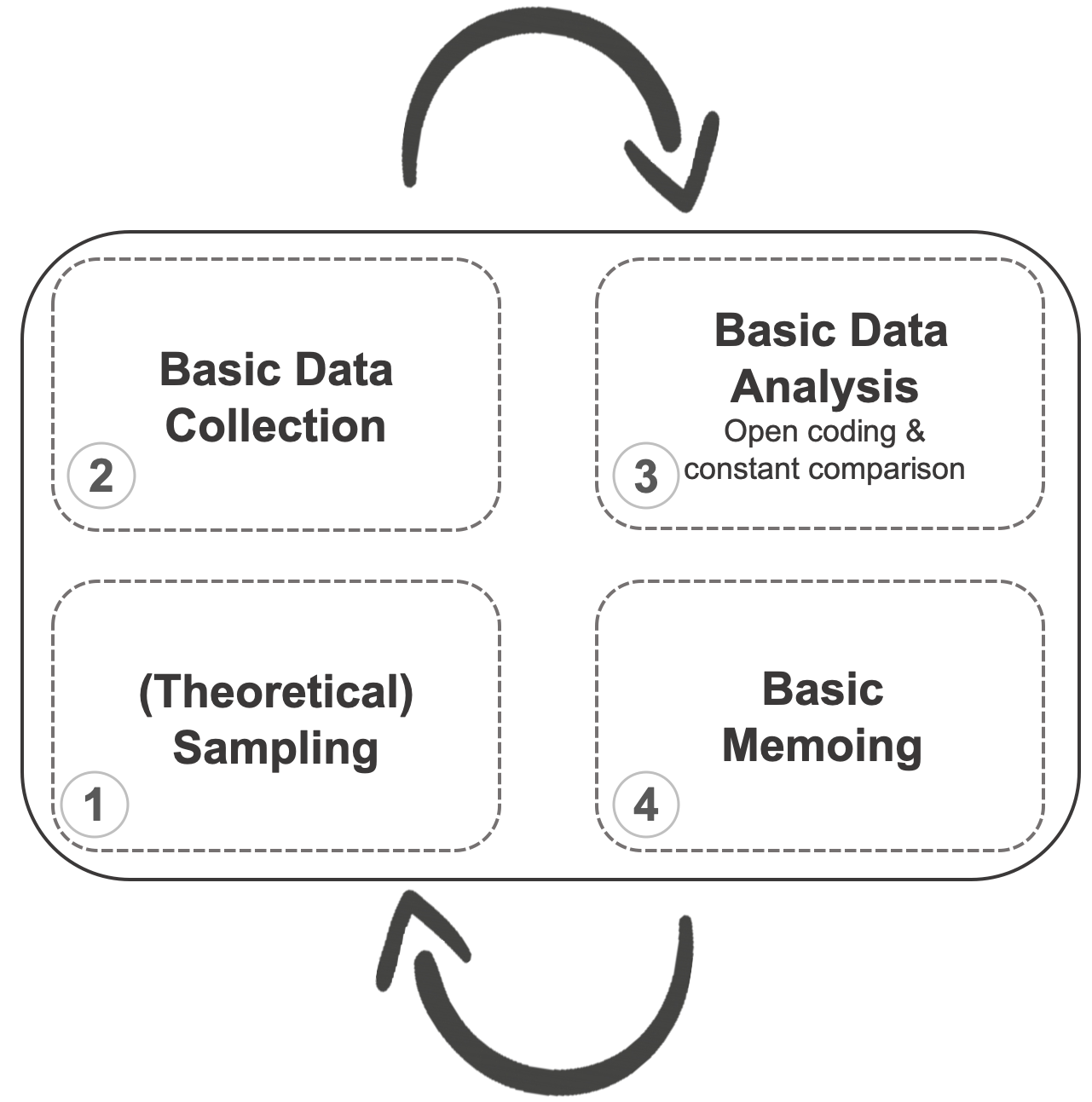}
\end{wrapfigure}

Basic data collection and analysis (DCA) is an iterative step involving the procedures of \textit{(theoretical) sampling}, \textit{basic data collection}, \textit{basic data analysis}, and \textit{basic memoing}. 

\subsubsection*{(Theoretical) Sampling}
STGT supports \textit{different sampling} techniques to get data collection underway, e.g. convenience, purposive, random, or representative sampling, as applicable to the project context and constraints. Once the iterations of basic data collection and analysis start yielding concepts and categories, \textit{theoretical sampling} can be employed.  Theoretical sampling is \textit{the ongoing process of assessing the emerging codes, concepts, (sub)categories, and hypotheses, and targeting specific data sources and types for collection in the upcoming iterations that are likely to help identify, develop, and saturate them while filling any theoretical gaps.}

 Theoretical sampling has been described a ``\textit{complex form of sampling}'' \citep{coyne1997sampling}. While not including theoretical sampling explicitly, traditional views classify all sampling in qualitative research as purposeful, intentionally selecting data sources based on their specific characteristics \citep{patton1990qualitative, sandelowski1995sample}. In this sense, theoretical sampling has been classified as purposeful \citep{coyne1997sampling}, driven by the emerging theory, while its dynamic nature can also lend itself to the snowballing strategy. A distinguishing feature of theoretical sampling is that it is not pre-determined, rather \textit{ongoing} and drives interleaved data collection and analysis.

Because it involves assessing high-level findings to direct data collection in specific instances (moving from general to specific), theoretical sampling can be said to apply \textit{deductive reasoning} (see Table \ref{tab:typesofreasoning})  \citep{becker1993common}. For example, applying theoretical sampling to the emergent self-organising roles on agile teams based on data collected from relatively new agile teams in the early stages of the study \citep{hoda2010organizing} led to identifying theoretical gaps around the maturity of the roles. The later parts of the study employed theoretical sampling to target data collection from mature agile teams leading to the subsequent mature and saturated definitions of these roles \citep{hoda2012self}. Theoretical sampling includes grooming and refinement of the research protocols from time to time, e.g. refining interview questions or keywords for mining public repositories to progressively focus on emerging concepts.

Being aware of the full socio-technical context of the study domain will mean researchers can apply sampling effectively to identify data sources such as participants (e.g. software developers, users, stakeholders) and documentation (e.g. app reviews, GitHub wikis) and apply appropriate data collection techniques such as physical or virtual interviews and observations and automated or manual mining of software repositories and online forums.

\subsubsection*{Basic Data Collection}
STGT accepts a variety of data sources as well as collection approaches.  The modern socio-technical world with its multiple realities offers new \textit{public sources} (e.g. publicly available practitioner blogs, presentations and talks as YouTube videos, and opinion trends on Twitter) and \textit{domain-specific sources} (e.g. digital project management boards such as Trello, open source software (OSS) repositories such as GitHub, communication channels such as Slack) and modern \textit{collection approaches} (e.g. immersive and experiential approaches on online social media or in virtual game worlds.)

Data collected from traditional sources such as semi-structured interviews and observations continue to be important, providing rich information, facial and voice cues, and opportunities for follow-up questioning. They offer real examples verifying participant statements and adding necessary contextual background, but are limited in scale due to the manual effort required.  Modern data publicly available, on the other hand, e.g. online forums and OSS repositories, and modern techniques such as data mining, offer unique opportunities to collect large amounts of qualitative data. However, these data are not custom elicited and will likely require more effort in selecting, filtering, and cleaning before the can be used for analysis. 

A combination of traditional and modern approaches is recommended to achieve best results, where applicable. For example, an STGT study can use traditional data collection techniques such as interviews to elicit an initial set of rich and highly relevant concepts and (sub)categories during the \textit{basic stage}. These can be used as \textit{seed terms} to perform large-scale targeted data collection on public and open source data. Without the initial traditional data collection, it would be challenging to identify, select, and filter from the large numbers of and massive public and open datasets. Without modern data sources and techniques, it would be nearly impossible to manually scale the theoretical outputs, e.g. to be more robust and include a wider set of applicable contexts. A combination of both offers unprecedented relevance, robustness, and scale in theory development.

The open nature of modern data sources, where anyone can add any information with minimal authentication and scrutiny but with mass reach, poses new challenges for researchers. While traditional Glaserian GT accepts ``\textit{all is data}'' \citep{glaser1967discovery}, STGT nuances the original stance with the necessary caution required to assess modern data sources and accepts \textit{credible} data.  It also highlights the importance of ethical considerations, especially associated with modern data collection and usage (upcoming in section 7.3). That is, for STGT studies, \textit{all is data, with care}.

\subsubsection*{Basic Data Analysis}
Basic data analysis includes \textit{open coding} and \textit{constant comparison}. \textit{Coding} is the process of representing textual raw data into condensed formats that best capture its essence and meaning. Using a socio-technical approach to coding enables the researcher to reach beyond analysing the social context and meaning. It ensures the technical aspects are neither ignored nor treated as a token or decorative element (e.g. like a \textit{black box}), rather they are considered together with the social aspects to capture a more comprehensive socio-technical essence and meaning of the raw data.

\textit{Open coding} refers to the coding applied in the early stages of the research study where the researcher remains \textit{open} to any and all codes arising, ensuring a comprehensive coverage. Because of its open nature, open coding is likely to result is large amounts of code. \textit{Constant comparison} is the process of constantly comparing derived codes within the same source and across sources to identify key patterns in the data. Constant comparison concretely manifests an \textit{inductive} approach, leading from specific instances toward general patterns, within the study context. Constant comparison is applied at increasing levels of abstraction to raise the codes to the concept level, concepts to sub-category (where applicable) and category levels. 

Application of a traditional sociological approach to GT data analysis in a software engineering research context is likely to identify social concepts and come up with rich theories comprising them.  For example, applying open coding the traditional way can lead to the identification of concepts that capture how software development teams regularly engage in social activities of \textit{brainstorming}, \textit{discussing}, \textit{debating}, and \textit{reconciling choices}. However, software engineering is not simply a mixture of social and technical aspects studied in isolation. Socio-technical data analysis in STGT refers to analysing the complete socio-technical context. For example, applying STGT in the same context is likely to uncover and explain the full socio-technical nature of the same activities as \textit{brainstorming software requirements}, \textit{discussing software architecture}, \textit{debating low fidelity prototypes}, and \textit{reconciling choices of diverse technologies and platforms}. These concepts encapsulate more comprehensive socio-technical meanings and essences of the activities being observed and are more likely to be relevant to and benefit practitioners, a key aim of GT studies. Researchers applying a socio-technical approach to open coding are likely to apply high levels of \textit{socio-technical sensitivity}, picking up on the nuances of different socio-technical practices involving seemingly similar social activities, e.g. \textit{debating low fidelity prototypes} as opposed to \textit{debating design patterns}. Where developers may be biased toward the prototypes they create themselves and therefore more defensive in their debating strategies in the former case and more open to logical argumentation when debating established design patterns in the latter case.

Interestingly, most quality GT studies in software engineering seem to `naturally' apply a socio-technical adaptation to data analysis, although rarely acknowledged and hitherto undefined and unexplained, leading to the formulation of socio-technical grounded theories, e.g., theories of \textit{reconciling agile and software architecture} \citep{waterman2015much}, \textit{becoming agile} \citep{hoda2017becoming}, and \textit{variations in scrum practice} \citep{masood2020tse}. More examples of applying a socio-technical approach to open coding can be found in \citep{hoda2011impact, hoda2012self, hoda2017becoming, masood2020tse, masood2020emse, yogi2021scrummaster, yogi2021jss}. 

 Explicitly acknowledging and explaining the data's socio-technical nature will enable aspiring researchers to conduct quality STGT data analysis. Additionally, while it was considered ``\textit{very shortsighted to believe that computers are capable of gleaning the meaning embedded in qualitative data}'' \citep{becker1993common}, rapid technological advancements in natural language processing, sentiment analysis, and other AI based analysis techniques, promise further easing and augmenting of manual analysis, enabling unprecedented improvements and scaling of qualitative data analysis and theory development in the future.

\subsubsection*{Basic Memoing}
\textit{Basic memoing} is the process of documenting the researcher's thoughts, ideas, and reflections on emerging concepts and (sub)categories and evidence-based conjectures on possible links between them. Memos can be in the form of written notes, sketches, annotated images or photographs. While voice and video recording can also be used, it is recommended these are transcribed for ease of further comparison and analysis in the advanced stage.

Memoing is an imperative procedure that distinguishes STGT from other qualitative research methods. It is the mechanism through which researcher reflection is systematically documented and used for theory development. Examples of memos can be found in \citep{hoda2012developing} and \citep{masood2020emse}.\\

\noindent The basic data analysis stage enables a progressive condensation of the large amounts of initial codes generated during open coding using constant comparison and memoing into concepts and categories, and narrows the focus of analysis. As key concepts, (sub)categories, and preliminary hypotheses start to emerge, the researcher is ready for the advanced theory development stage. Practically, this is manifested in a few key categories being strong and detailed enough to be presented for peer-review, and subsequent workshop presentations, and conference or journal publications.

\section{STGT Method -- Advanced Theory Development}
\label{advancedtheorydev}

STGT acknowledges the challenges of the novice socio-technical researcher in understanding each of the traditional sociological GT versions (\textit{Glaserian}, \textit{Strauss-Corbinian}, \textit{Constructivist}), differences between them, and selecting one version upfront with no room for an easy switch or combining of procedures from different versions later on. STGT delays the decision making to an advanced stage where the researcher has experienced basic data analysis procedures and is in a position to understand and decide which mode of advanced data analysis and theory development best applies to their emergent findings. 

 In the advanced stage, STGT offers a choice of two modes of theory development, \textit{emergent} and \textit{structured}, resulting in mature, structured and integrated theories as outcomes. Table \ref{tab:theorydevmodes} provides an overview of their differences and commonalities. Both modes include \textit{targeted literature review} and \textit{theoretical sampling}, as described earlier. Both modes also have \textit{advanced memoing} and \textit{theoretical saturation} in common, described next.

\subsubsection*{Advanced Memoing}
Advanced memoing is common to both emergent and structured modes of theory development and involves careful revisiting, reassessing, refining, and comparing of memos. Advanced memoing\footnote{Advanced memoing proposed by STGT replaces \textit{theoretical sorting} in \textit{ Glaserian} and \textit{Strauss-Corbinian} GT \citep{glaser1967discovery, strauss1990basics}, acknowledging hypotheses development and structuring as iterative and incremental rather than a one-time activity.} serves to confirm and strengthen existing relationships, reject unsupported indicative relationships, and establish new links between (sub)categories.

Reflections captured in memos help develop and solidify emerging concepts and (sub)categories. When memos are compared and related to one another, insights in the form of possible links between different concepts and (sub)categories start to emerge. Exploring these potential links (aka hypotheses) requires \textit{abductive reasoning} -- proposing the best possible explanation based on all available evidence (see Table \ref{tab:typesofreasoning} -- \textit{types of reasoning}). As more evidence is collected, it adds to the strength of the proposed explanations and emerging links. Applying a socio-technical approach to memoing ensures that the underlying data encapsulates the full socio-technical research context and the researcher can propose explanations based on socio-technical knowledge and understanding. For example, a memo may capture the varying levels of scrum master involvement in team practices across multiple teams observed and propose a relationship with the agile maturity of the team where novice teams show more scrum master involvement as compared to experienced agile teams. Such hypothesis development requires not only an understanding of the social relationships between the scrum master and the team but also an understanding of the team's socio-technical practices, e.g. daily standup and pair programming, to grasp the level of scrum master involvement to be expected as compared to what is observed.

\begin{table}[]
    \centering
    \caption{\small Emergent vs structured theory development}
    \scriptsize
    \begin{tabular}{>{\raggedright\arraybackslash}p{0.8cm}>{\raggedright\arraybackslash}p{3.2cm}>{\raggedright\arraybackslash}p{3.2cm}}
    \hline\noalign{\smallskip}
    & \textbf{Emergent mode} & \textbf{Structured mode}\\
    \hline\noalign{\smallskip}
    Context of Use 
    & Broad phenomenon (e.g. \textit{exploring agile project management}). & Narrow phenomenon (e.g. \textit{investigating self-assignment in practice}).\\
    \noalign{\smallskip}
    & Open coding applied widely on broad phenomemon. & Open coding applied on narrow phenomenon.\\
    & Basic stage ended with the emergence of some strong and some not so strong (sub)categories and indicative relationships. & Basic stage ended with the emergence of clear set of (sub)categories that form an overarching structure.\\
    \noalign{\smallskip}
    Distinct Steps 
    & Iterations of \textit{targeted} data collection and analysis, targeting most significant (sub)categories and continuing to establish relationships between them. & Iterations of \textit{structured} data collection and analysis, focusing on saturating individual (sub)categories and firming up relationships between them.\\
    \noalign{\smallskip}
    & Theoretical structuring. & Theoretical integration.\\
    \end{tabular}
    \label{tab:theorydevmodes}
\end{table}
\begin{table}[]
    \centering
    \scriptsize
    \begin{tabular}{p{0.8cm} p{7cm}}
    \hline\noalign{\smallskip}
    Common Steps & Targeted literature review, theoretical sampling, advanced memoing, theoretical saturation.\\
    \noalign{\smallskip}
    Outcome & Structured and integrated theory or theories. \\
    \noalign{\smallskip}\hline\noalign{\smallskip}
    \end{tabular}
\end{table}

\subsubsection*{Theoretical Saturation} 
When data collection reaches a point of diminishing returns where further collection does not generate new or significantly add to existing concepts, categories, or insights, the study has reached \textit{theoretical saturation}. Practically, this manifested as the last few data collection attempts (e.g. last 3-4 interviews) serving to verify the theory. The theory is considered mature as it comprises of strongly supported and dense categories and hypotheses.

\subsection{Emergent Mode of Theory Development (Option 1)}
The emergent mode allows the researcher to conduct further rounds of iterative and interleaved data collection and analysis in a \textit{targeted} manner. The emergent mode enables an organic emergence of theory as recommended by Glaserian GT \citep{glaser1967discovery}. The emergent mode includes iterations of \textit{targeted} data collection and analysis ending with theoretical saturation and followed by \textit{theoretical structuring} to derive mature theory.

\subsubsection*{ Context of Use}
 Where the initial research phenomenon or topic was broad (e.g. \textit{exploring agile project management} \citep{hoda2012developing}) and open coding was applied extensively on the broad level, the basic stage will likely lead to the emergence of a number of (sub)categories, some stronger than others, with indicative relationships emerging between them. In this case, where a some strong categories and initial relationships have emerged but a clear theoretical structure is not evident, the researcher can decide to proceed with an emergent mode of theory development which relies on the theoretical structure emerging progressively over time.
    
\subsubsection*{Targeted Data Collection} 
Data collection in the emergent mode is driven by \textit{theoretical sampling} as described earlier, targeting data sources most likely to help fill theoretical gaps and strengthen key categories, and including refinements of the data collection protocols to progressively focus on the key categories.\\
    
\subsubsection*{Targeted Data Analysis}
\begin{wrapfigure}[13]{l}{0.25\textwidth}
    \centering
    \vspace{-9pt}
    \includegraphics[width=0.25\textwidth]{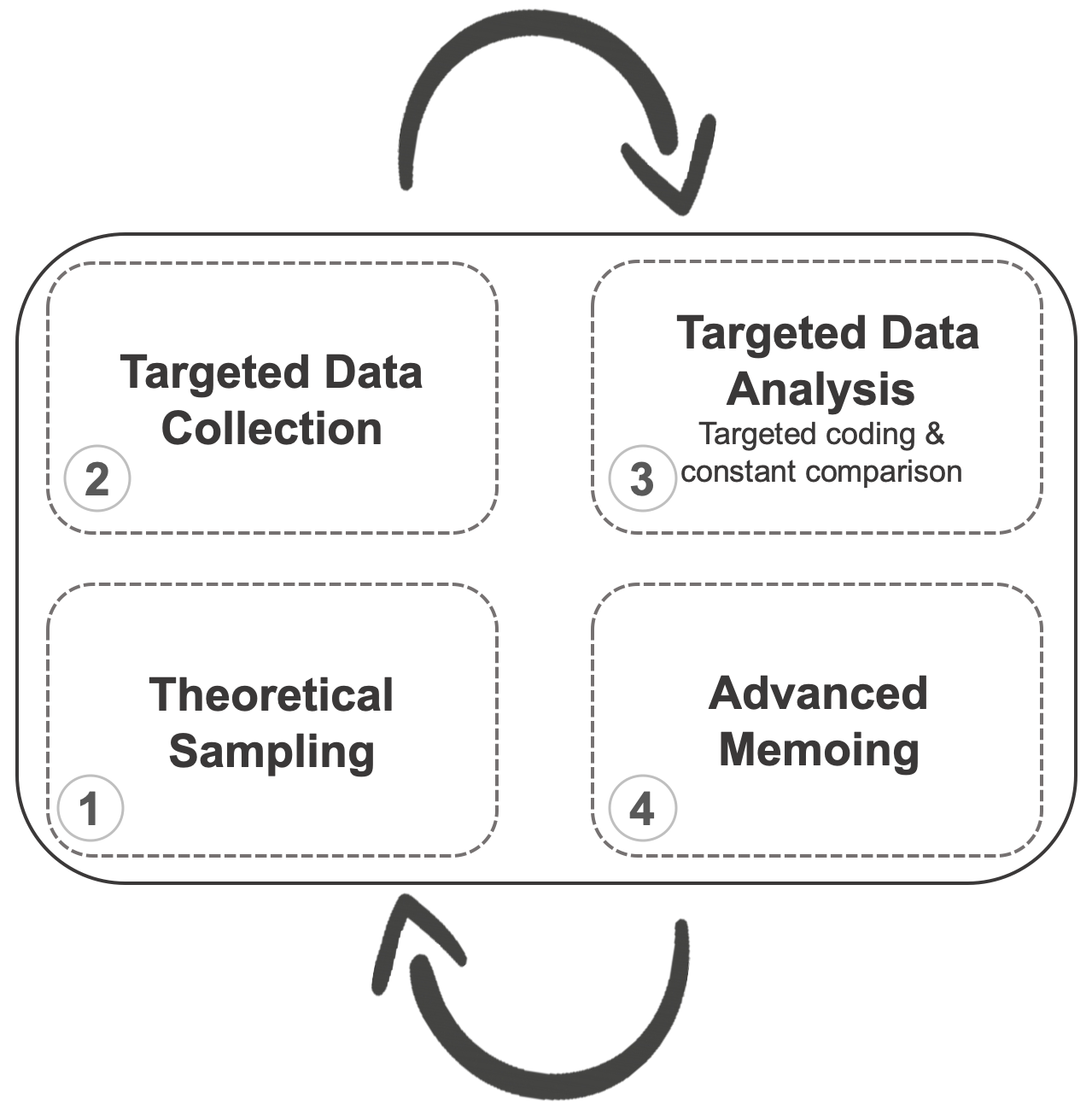}
\end{wrapfigure} 
Targeted data analysis involves \textit{targeted coding} which is the same procedure as open coding but applied in a targeted manner, and \textit{constant comparison}. Unlike open coding, targeted coding\footnote{Targeted coding is the same as selective coding in \textit{Glaserian GT} but STGT avoids using the term selective coding since it refers to two very different procedures in \textit{Glaserian GT} and \textit{Strauss-Corbinian GT}. To avoid confusion, STGT uses the term \textit{targeted} coding.} limits analysis to only the most significant concepts and categories arising from the basic stage. However, some new categories can still emerge during targeted data analysis. Advanced memoing ensures the links between (sub)categories are strengthened.
    
\subsubsection*{Theoretical Structuring} 
Once \textit{theoretical saturation} is achieved, the researcher can decide to present the theory as is, using the emergent structure. However, STGT recommends the researcher to perform \textit{theoretical structuring} by (a) exploring and identifying the genre of theories that best fits, e.g. process, taxonomy, degree, strategies \citep{glaser1978theoretical}, and (b) exploring if the emergent theoretical structure naturally maps to a pre-existing theoretical template, e.g. six C's model, process template with defined stages and entry/exit indicators, and in case of a good fit, use the template to further structure their theory. Theoretical structuring\footnote{Theoretical structuring in essence is the same as \textit{theoretical coding} in Glaserian GT but STGT avoids using that term since the procedure does not involve any actual \textit{coding}, rather structuring or organising of the emergent theory.} is an optional but highly recommended procedure. 

The \textit{theory of becoming agile} is an example of presenting a mature theory in its emergent structure, i.e. a bespoke process model \citep{hoda2017becoming}. The theory was classified under the genre of \textit{process theory} since it explained the \textit{process} of software teams becoming agile teams but no explicit pre-defined structure was adopted. On the other hand, in case of the \textit{theory of inadequate customer collaboration} \citep{hoda2011impact}, the pre-defined \textit{six C's theoretical template} was adopted at a late stage as a result of exploring Glaser's theoretical coding families and discovering a good fit with the six C's template \citep{glaser1978theoretical}.

\subsection{Structured Mode of Theory Development (Option 2)}

The structured mode allows the researcher to conduct further rounds of iterative and interleaved data collection and analysis in a \textit{structured} manner. The structured mode enables structured development of theory as captured by the early forms of \textit{Strauss-Corbinian} GT \citep{strauss1990basics}. The structured mode includes iterations of \textit{structured data collection and analysis} ending with theoretical saturation and followed by \textit{theoretical integration} to derive mature theory.

In the structured mode, the researcher can explore possible fits with pre-defined theoretical templates early, coming out of the basic stage. Where fitting, a pre-defined theoretical template can be adopted and used to guide the remaining data collection, analysis, and theory development. For example, templates such as the \textit{coding paradigm} (e.g. \citep{masood2020emse}), \textit{conditional matrix} \citep{strauss1990basics}, or the Glaserian six C's template can be used. Alternatively, the researcher can use their own theoretical structure emerging out of the basic stage to guide further data collection, analysis, and theory development, e.g. \citep{masood2020tse}). In either case, the advanced stage is guided by a theoretical structure already in place.  \cite{sjoberg2008theory}'s guidance on defining constructs and propositions provides an early example of using a ``\textit{grounded theory}'' based approach to structured theory development within an exploratory case study.

\subsubsection*{ Context of Use}
 Where the initial research topic was relatively narrow and well defined (e.g. \textit{investigating self-assignment practice} \citep{masood2020emse}) and the data collection was focused on well defined facets of the topic, the basic stage can lead to the emergence of a set of (sub)categories and a relatively clear theoretical structure. In this case, it makes sense for the researcher to proceed with a structured mode of theory development.\\

\subsubsection*{Structured Data Collection} 
Data collection in the structured mode is driven by \textit{theoretical sampling}, collecting data to progressively fit, add, strengthen, and saturate the concepts and (sub)categories already present in the guiding theoretical structure. It includes refinements of the data collection protocols to progressively saturate existing categories and occasionally identify new concepts that appear within the theoretical structure.

\subsubsection*{Structured Data Analysis}
\begin{wrapfigure}[13]{l}{0.25\textwidth}
    \centering
    \vspace{-9pt}
    \includegraphics[width=0.25\textwidth]{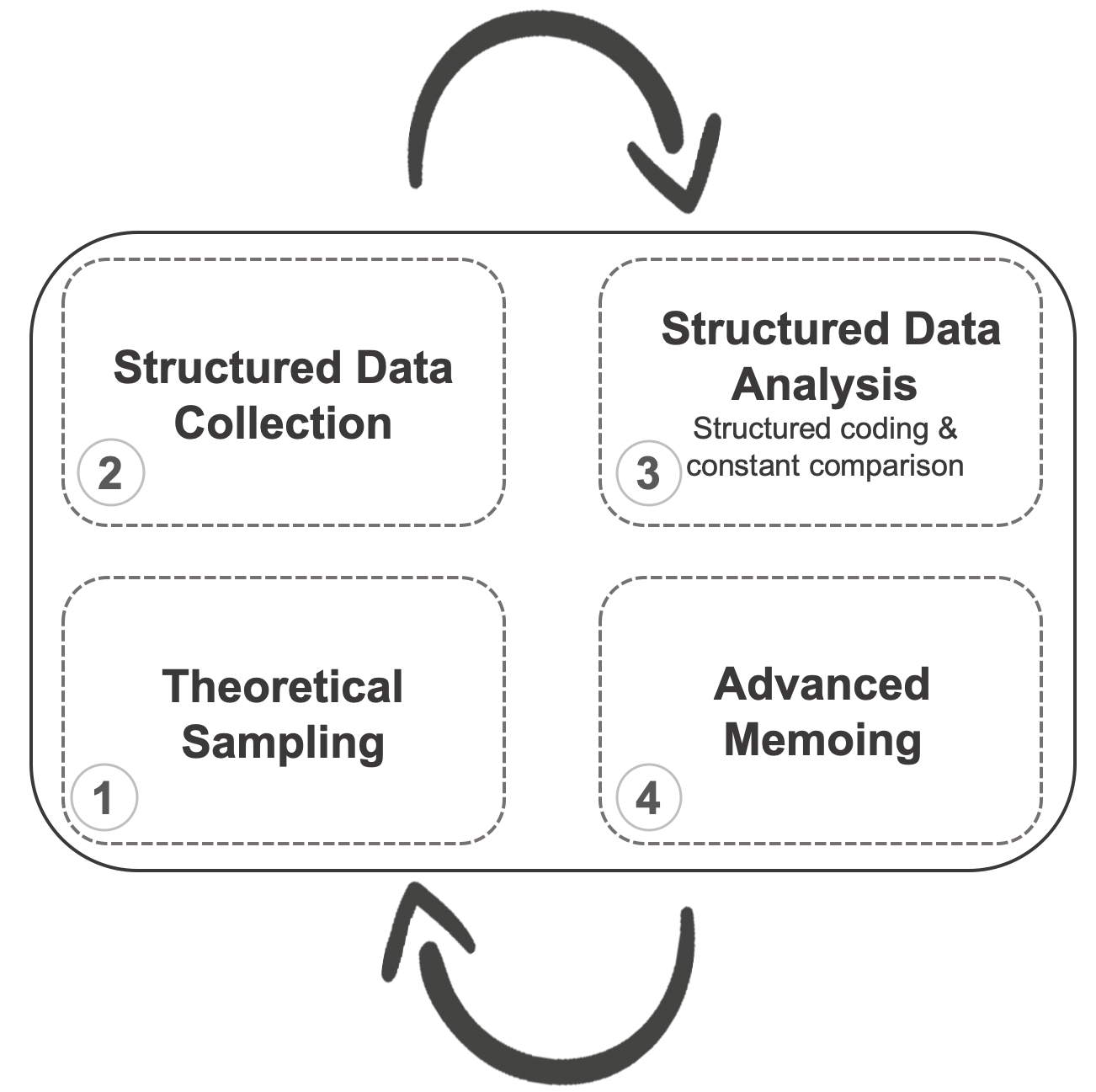}
\end{wrapfigure} Data analysis in the structured mode involves \textit{structured coding} and \textit{constant comparison}. Structured coding is similar to the original \textit{axial coding} introduced in \citep{strauss1990basics} in that it applies data analysis using a guiding theoretical structure and aims to identify and strengthen the relationships between the key categories. Advanced \textit{memoing} ensures the links between (sub)categories are strengthened.

\begin{figure*}
    \centering
    \includegraphics[width=0.99\textwidth]{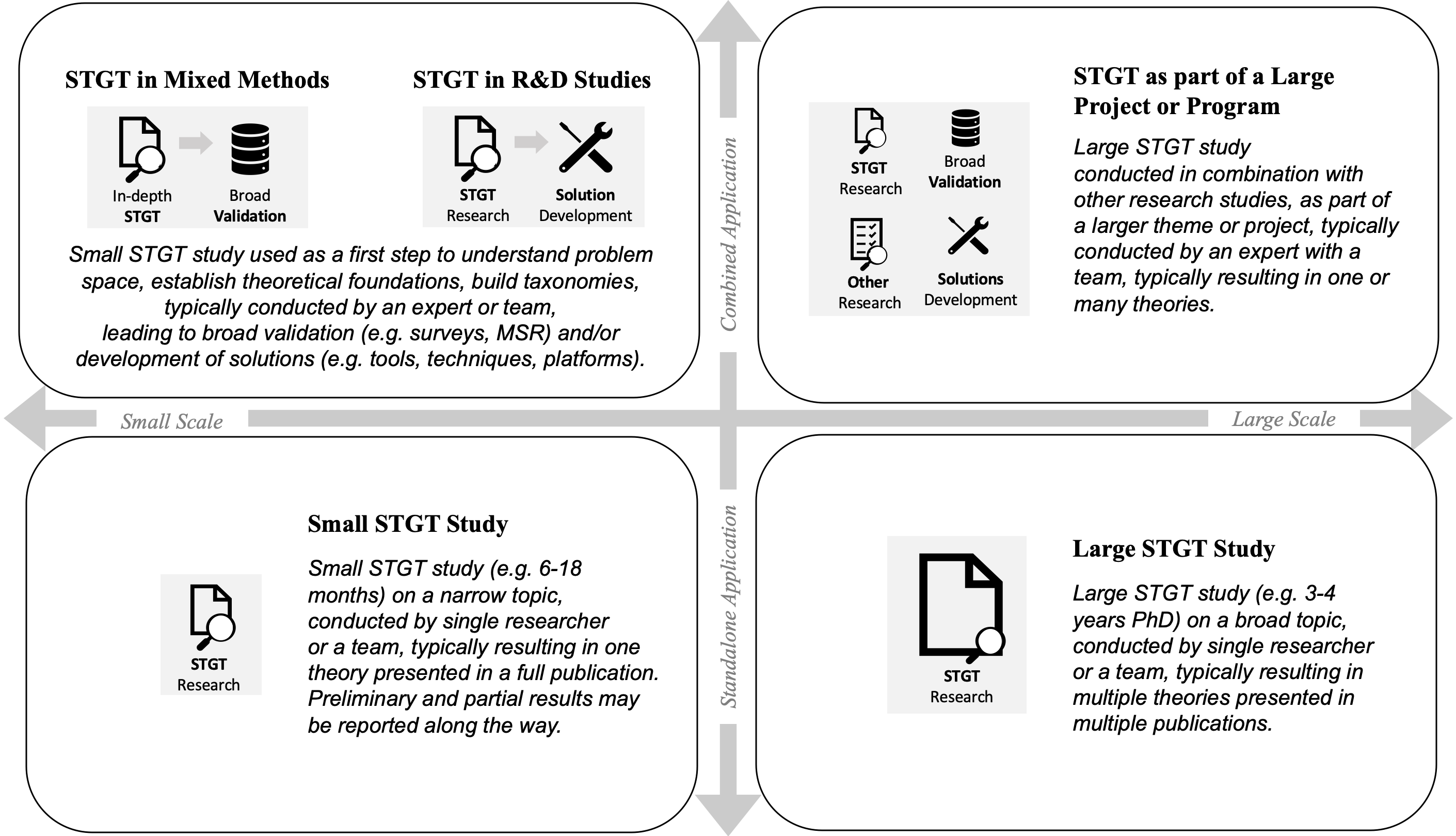}
    \caption{Range of STGT Applications  (clockwise from right): STGT in mixed methods, STGT in R\&D studies, STGT as part of large projects or programs, large STGT study, small STGT study. MSR stands for mining software repositories.}
    \label{fig:STGTApplications}
\end{figure*}

\subsubsection*{Theoretical integration} 

Once \textit{theoretical saturation} is reached in the structured mode, the theory should comprise of dense and strongly supported categories and hypotheses and is considered both structured and mature. But, it may not be fully integrated. Theoretical integration is the procedure of ensuring the (sub)categories are fully integrated into the overall theoretical structure. Practically, this involves finalising the \textit{story-line} of the theory, answering the question, \textit{what is this theory about?} or \textit{what phenomenon is captured and explained by this theory?} For example, in case of \citep{masood2020emse}, theoretical integration toward the later stages involved discussing what basic phenomenon the categories and relationships were best capturing. \textit{How agile teams make self-assignment work} was decided as the overall story-line or phenomenon that the categories and relationships best helped explain.

Sometimes, integration can result in the emergence of an additional theoretical layer that serves to support and enhance the existing structure. For example, the \textit{theory of Scrum variations} was structured around three categories, variations in Scrum roles, variations in Scrum practices, and variations in Scrum artefacts \citep{masood2020tse}. In the final stages of integrating the theory, a \textit{nuanced classification system} emerged which served to further classify the variations across the three categories in terms of \textit{standard, necessary,} and \textit{contextual variations}, and \textit{clear deviations}.

\section{STGT - Application, Outcomes, Reporting}
\label{application}

Socio-technical grounded theory can be applied in two ways. First, as a \textbf{full STGT method} resulting in novel, useful, parsimonious, and modifiable \textit{\textbf{theories}} grounded in evidence. And second, in a limited capacity as \textbf{STGT for data analysis}, by applying its basic data analysis techniques, within or alongside other research methods, producing outcomes of original and relevant \textit{\textbf{descriptive findings}} in the form of key categories or preliminary hypotheses and propositions. Figure \ref{fig:STGTApplications} shows types of STGT applications and associated possible outcomes. It is important that researchers understand, select, and clearly present their application approach and claim the appropriate outcomes and reviewers evaluate relevant outcomes accordingly, using the STGT Evaluation Guidelines (section \ref{evaluation}).

\subsection{Full STGT Method}

Full STGT method refers to the application of the wider STGT method guidelines beyond basic data collection and analysis and including advanced stages of theory development, ultimately resulting in mature theories (see Figure \ref{fig:STGT}). Full applications can be both \textbf{standalone} and as part of \textbf{combined} studies within larger mixed methods frameworks that apply other research methods, e.g. action research or survey research, and/or are conducted as the \textit{research} component of research and development (R{\&}D) projects.

\subsubsection*{Standalone STGT Studies}
Standalone STGT studies refer to a full STGT method application in a standalone capacity and do not apply other research methods such as  survey research to further validate the theory. Mixed data, both qualitative and quantitative data, can be used within standalone studies. Standalone applications result in what \cite{storey2020EMSE} refer to as \textit{descriptive knowledge}, with potentially limited \textit{prescriptive knowledge} in the form of practical guidelines and recommendations (e.g. as seen in \cite{masood2020emse}, \cite{masood2020tse}, and \cite{yogi2021scrummaster}.) 

Standalone STGT studies can vary in scale. They can take the form of small studies that address narrow research topics, e.g. \textit{impact of human values on software design decisions}, \textit{mob programming techniques}, or \textit{the use of coupling and cohesion in microservices design}, the last derived from a Twitter post by Grady Booch. They are conducted over a relatively small duration of time, e.g. six to eighteen months, typically by a single researcher or a team of researchers. Small standalone STGT studies are likely to result in one mature theory, usually presented in a single publication, although preliminary and partial results can be reported along the way. 

They can also take the form of large scale studies that address broad research topics, e.g. \textit{role of ethics in artificial intelligence}, \textit{future of work in a post pandemic world}, and \textit{digital equity for the elderly}. A comprehensive treatment of such topics requires several years, typically by a team of researchers. A standalone STGT study conducted as a part of a PhD program, typically three to four years in length, is an example of a large-scale STGT study. These are likely to result in multiple theories, usually published as multiple articles. How the multiple theories `\textit{hang together}' is usually detailed in the thesis report and should be briefly explained in individual articles with cross-reference to related articles.

\subsubsection*{Combined STGT Studies}
There is an increasing trend of performing mixed methods, combined studies, and R{\&}D projects in SE. In such cases, a full STGT study serves as a robust mechanism to \textit{research} the problem space, leading to the \textit{development} of highly relevant and customised solutions. For example, a small scale STGT study can be conducted as a first step to establish theoretical foundations, explain the problem in depth, and/or devise taxonomies followed by wider validation and/or tools development. Outcomes in these cases are expected to be what \cite{storey2020EMSE} refer to as \textit{descriptive knowledge} from the STGT research part that then ideally feed into or lead to the development of \textit{solutions}, e.g. tools, techniques, platforms, from the application of other methods or in the development component of the R{\&}D project.

\subsection{STGT for Data Analysis}
A limited application of STGT is possible through using its \textbf{basic data analysis procedures} of \textit{open coding} and \textit{constant comparison}, including \textit{memoing}, within a different research method or framework.  For example, the qualitative data collected as part of a case study, survey study, ethnography, action research, interview-based study, or even experiments, can be analysed using STGT's basic data analysis procedures. These are likely to produce \textit{descriptive findings} that encapsulate and describe dense codes, concepts, and categories. Using memoing, the beginnings of relationships between the concepts, and (sub)categories may become evident. These can be presented as \textit{preliminary hypotheses} or \textit{propositions} (e.g. use of GT for data analysis within a case study \citep{bick2017coordination} or within a mixed method research study \citep{madampe2020towards}). 

In the absence of applying the remaining STGT steps and advanced data analysis and theory development procedures, the outcomes cannot and should not be claimed as a mature theory. Why is this so? Where both STGT's basic and advanced data analysis procedures are applied within other methods, the study enters a murky area where it becomes difficult to justify outcomes as grounded theories at par with those produced through the application of the full STGT method propelled by theoretical sampling and iterative and interleaved data collection and analysis. In such cases, reviewers may question the motivations for the mixed approach along with asking why the full STGT method was not applied given its core data analysis steps -- often regarded as the most rigorous part -- were applied. In these cases, researchers are strongly recommended to consider applying the full STGT method.

\subsection{Ethical Concerns}

Conducting STGT studies relies on collecting and analysing data, publicly available and custom collected for the study. An early ethical assessment should be conducted to consider potential ethical concerns with the type and means of data collection and reporting. Where applicable, approval should be gained from the relevant ethical committee. 

Given the wide range of data sources and techniques STGT studies can leverage, it is critical that ethical concerns are explicitly addressed. For example, for data that is custom collected with the informed consent of participants -- such as through interviews, focus groups, observations, surveys, questionnaires -- conditions of participant confidentiality and anonymity can be maintained by obscuring the identifiable details and sharing selected sanitised parts of the underlying data when reporting findings. 

However, this is challenging when working with public data, typically available through modern, digital sources such as project repositories, social media posts, and YouTube videos. Where possible, STGT researchers should seek explicit consent (e.g. requesting a speaker to use their interview or talk video on YouTube for research purposes) or obscure identifiable information, especially in case of potential harm arising from reported research (e.g. a study using public data as examples of `bad'  programming practices with traceable individual or organisational identities). 

A number of ethics resources can be consulted. For example, \citep{EthicsTSE} and the articles contained in a special issue on research ethics \citep{EthicsEmpiricalSE} provide numerous examples of ethical issues and practice recommendations specific to SE research such as \textit{informed consent}, \textit{beneficence to individual and organisation} (maximizing benefits and minimizing harm to society and participants), and \textit{confidentiality}. Guidelines on ethical elicitation and responsible use of public data, e.g. social media ethics framework \citep{EthicsSocialMedia}, ethical decision-making in internet research \citep{EthicsInternet} and more are also available. The Menlo \citep{EthicsMenloReport} report lays down the ethical principles guiding information and communication technology research  as \textit{respect for persons}, \textit{beneficence}, \textit{justice}, \textit{respect for law}, and \textit{public interest}. It also shares helpful approaches to identifying and balancing potential benefits and risks. \cite{EthicsMSR} used the Menlo framework to review and discuss the ethical implications of mining software repository (MSR) research across sources such as IDE events, version control data, build logs, stack overflow, issue trackers, and mailing lists. They identified a need for open discussion of ethical issues in the MSR community. Similar efforts have been made in the open source community, e.g. \citep{EthicsOS}. STGT researchers should consult and apply ethical research frameworks and guides applicable to the data sources, types, and collection techniques they intend to use.

\subsection{Reporting STGT}
Where a researcher has applied the full STGT method rigorously and produced outcomes in the form of dense categories and supported relationships between them, they can and should claim and present it as a \textit{mature theory}. The mature theory, whether developed using an emergent or structured mode, is ready for reporting. Because of the size and density of a mature theory, it is best presented in a journal, thesis, or book format which allows for enough room to explain all its constituent parts in detail. However, a mature theory can also be presented in a shorter format such as a conference paper in a highly condensed form with an overarching view of the phenomenon (e.g. \textit{theory of becoming agile} \citep{hoda2017becoming}).

Feedback on emerging concepts, categories, and hypotheses is also important, both from the practitioner and research communities to assess and improve relevance and rigour respectively. STGT researchers can submit \textit{partial results} in the form of an emerging or saturated (sub)category or categories, or \textit{preliminary theory} to research workshops, conferences, and journals. In doing so, they should note the paper reports on a specific (sub)category or \textit{preliminary} theory. Where a number of papers are prepared resulting from the same underlying study, which can easily happen with large STGT studies, they should present a high level overview of the larger study and how the different papers relate to each other. Similarly, the overarching STGT application, e.g. as a full STGT study, should be explained while clarifying parts being reported. STGT researchers are strongly recommended to present their emerging findings to practitioner groups, including both their participants and others not related to the study, to receive feedback on how well they resonate with practice. Peer review and practitioner feedback help identify theoretical gaps and can further guide theory development.

\section{STGT -- Evaluation Guidelines}
\label{evaluation}
Evaluating an STGT manuscript (report, article, or thesis) involves evaluating both the \textit{method application} and \textit{outcomes}. Below I present a set of criteria for evaluating STGT studies, including applications and outcomes. While general quality criteria expected of most research studies apply, I focus on specific criteria that apply to STGT studies. These do not aim to constitute a \textit{checklist approach} and are deliberately \textit{nuanced} to distinguish between method application and outcomes, and further between preliminary (or partial) and mature outcomes. Authors can use these as guidelines to ensure they present strong manuscripts and reviewers can use these as evaluation criteria. Table \ref{tab:evaluation} summarises the evaluation criteria for the method application and outcomes.

\begin{table}[t]
    \centering
    \caption{\small STGT Evaluation Criteria}
   \begin{tabular}{>{\raggedright\arraybackslash}p{2.2cm}>{\raggedright\arraybackslash}p{2.6cm}>{\raggedright\arraybackslash}p{2.4cm}}
    \hline\noalign{\smallskip}
    \textbf{Method Application} & \textbf{Partial findings or Preliminary theory} & \textbf{Mature theory}\\
    \hline\noalign{\smallskip}
   Credibility, Rigour & Originality, Relevance, \textcolor{white}{space} Density & Novelty, Usefulness, Parsimony, Modifiability\\ 
    \hline\noalign{\smallskip}
    \end{tabular}
    \label{tab:evaluation}
\end{table}

\subsection{Evaluating STGT Method Application}
The research method is the bridge that connects the study aims to its outcomes. When a manuscript is submitted for review, it invites the reviewers to cross that bridge. If the methodology bridge is assessed as shaky or unreliable, it will raise concerns about the robustness of the study and trustworthiness of its outcomes. Conversely, exemplar studies include sufficient method application details (see Table \ref{tab:ExemplarGTstudies}). A well done STGT study will display and can be evaluated on two overarching criteria -- \textit{credibility} and \textit{rigour}.\\

\noindent\textbullet\space\space  \textbf{Credibility.}\\
Method application details provide evidence of clear understanding and effective application. Referring to or restating method guidelines is not enough. How those guidelines were practically applied in the study should be described, preferably including examples of application, to establish credibility. 
\vspace{-0.08cm}
\begin{itemize}
    \item[--] How were participants recruited? 
    \item[--] What initial sampling technique was applied?  
    \item[--] How was iterative and interleaved data collection and analysis ensured? 
    \item[--] How were memos written and used? 
    \end{itemize}
    
\noindent For manuscripts presenting mature theories, additional information should be provided.
\vspace{-0.08cm}
    \begin{itemize}
    \item[--] How was theoretical sampling applied?
    \item[--] How were the research protocols refined through the iterations? 
    \item[--] Which mode of theory development was applied, \textit{emergent} or \textit{structured}? 
    \item[--] How was theoretical saturation achieved?
    \item[--] What ontological and epistemological stances were used and why?
\end{itemize}
    
\noindent Depending on the ontological and epistemological stances, credibility may be assessed through additional criteria, e.g. replicability and reproducibility for positivist studies.

Lack of space in the manuscript is often cited as a justification for missing application details. Prioritising and treating the research methodology section as a \textit{first class citizen} in the manuscript will enable enough room to discuss these aspects. Alternatively, supplementary materials can be used to add to the basic details covered in the manuscript.\\

\noindent\textbullet\space\space  \textbf{Rigour.}\\
Sharing sufficient evidence of the underlying raw data, the data analysis procedures, and research artefacts, including evidence of theory development where applicable, within the manuscript is both important and possible. Providing enough and strong evidence helps establish both credibility and rigour of application.
\begin{itemize}
    \item[--] \textit{Example of basic coding}. The methodology section should include at least one \textit{working example} of the basic data analysis stage, covering open coding and constant comparison, that demonstrates how raw data was analysed to produce codes, concepts, and (sub)category (or categories, if presenting more than one category.)
    \item[--] \textit{Embedding of sanitised evidence}. Pertinent quotations and excerpts from the raw underlying data should be carefully interwoven with the description of the findings, ensuring all identifying details are removed or replaced with pseudo-names, to improve strength of evidence and add contextual depth and flavour. Similarly, photos from observations can be included after obfuscating faces and other identifying details.
    \item[--] \textit{Evidence of theory development}. For manuscripts presenting mature theories, evidence of conducting the advanced  theory development stage in \textit{emergent} or \textit{structured} mode should be supplied. For example, which coding structure was employed, where applicable.
\end{itemize}
    
Confidentiality and privacy concerns are often cited as justifications for not sharing evidence in qualitative studies. While these are very real ethical concerns, it does not equate to providing \textit{no evidence} or \textit{weak evidence}. Between complete disclosure that is impractical for qualitative studies and complete withholding of evidence, there is a balanced approach to establishing robustness through strength of evidence within the manuscript as described above. Evidence of memoing, additional working examples of the basic and advanced data analysis stages, descriptions of how a theoretical structure emerged or was applied can be included in supplementary materials.

\subsection{Evaluating STGT Outcomes}
STGT outcomes vary in nature. \textit{Partial findings} (e.g. one category being reported in-depth) or \textit{preliminary theories} (e.g. emerging hypotheses, propositions) result from the basic stage of \textit{a full STGT study} or the limited application of \textit{STGT for data analysis} with other methods. \textit{Mature theories} result as final outcomes of full STGT studies. Different outcomes cannot be evaluated using a \textit{one-size-fits-all} checklist.\\ 
    
\noindent \textbf{Partial outcomes} and \textbf{preliminary theories} should exhibit \textit{originality}, \textit{relevance}, and \textit{density}.\\

\noindent\textbullet\space\space  \textbf{Originality}.\\
Because an STGT study is purpose designed and executed, it is highly likely to produce original outcomes. Where originality is questioned, say because of how the outcomes map to or resemble existing research theories or practitioner models, originality should be demonstrated through:  (a) evidence of a purpose-designed and executed STGT study and practical steps taken to avoid biases arising from personal researcher backgrounds and explaining possible interplay between the resulting theory and related practitioner literature and pre-existing theories, in case of a positivist stance or (b) discussion of potential biases and particular perspectives and their role in the construction of the theory, in case of a constructivist approach.\\ 
    
\noindent\textbullet\space\space  \textbf{Relevance}.\\
Because an STGT study is empirically based, its findings should have relevance in same or similar contexts. Relevance is said to be achieved when feedback from participants and other practitioners serve to validate the findings. Reviewers can also independently assess relevance based on their wider knowledge of the domain. A side-effect of achieving relevance is that the findings can come across as ``\textit{unsurprising}'' or ``\textit{intuitive}'' to the experienced reader. A robust STGT application will ensure the findings, albeit intuitive, are never trivial or lacking depth.\\

\noindent\textbullet\space\space  \textbf{Density}.\\
Density refers to the depth or richness of categories. Density is said to be achieved when a category is supported by multiple concepts and properties that capture a range of contexts and nuances. The descriptions of the categories include multiple pertinent examples and evidence from the underlying data. Categories derived from the application of STGT are generally denser than  from descriptive or thematic analysis, reflecting the rigour of its data analysis procedures.\\

\noindent \textbf{Mature theories} resulting from full STGT studies need to be evaluated to higher standards, achieving the traits of \textit{novelty} (beyond originality), \textit{usefulness} (beyond relevance), \textit{parsimony} (beyond density), and \textit{modifiability}. \\

\noindent\textbullet\space\space  \textbf{Novelty.}\\ 
Since the main aim of STGT is to develop new theories instead of verifying existing ones, novelty is a key criterion for evaluating a mature theory outcome. Novelty is said to be achieved when the theory presents unprecedented insights on a relatively under-researched phenomenon (e.g. \citep{masood2020emse}) or a fresh approach to understanding a well-studied phenomenon (e.g. \citep{hoda2017becoming}). In fact, novelty can be assumed as a default once credibility and rigour of application are established. If a theory is seen to have close resemblance to existing models from research or practitioner sources, its novelty and the role of relevant models/literature can be further scrutinised. \\

\noindent\textbullet\space\space  \textbf{Usefulness}.\\
Mature theories go beyond being relevant and resonating with practice. They should provide actionable insights and recommendations for practice and preferably for research. Usefulness is said to be achieved when the  possible applications and implications for research and practice are shared, e.g. \citep{hoda2017becoming, masood2020emse}\\

\noindent\textbullet\space\space  \textbf{Parsimony}.\\
Parsimony refers to the compactness of the theory such that it exhibits conceptual density and explains potentially complex phenomenon in simple and elegant ways. Parsimony is said to be achieved when the theory can be described in a compact way, in a few sentences or a succinct paragraph that captures all the key categories and the relationships between them (e.g. see \textit{theory of Scrum variations} \citep{masood2020tse}), and preferably depicted in an easy to understand visual format, e.g. \citep{hoda2011impact, hoda2017becoming, masood2020emse} while its full description demonstrates rich, nuanced, and often multi-faceted or multi-layered findings that are otherwise difficult to achieve with other methods.\\

\noindent\textbullet\space\space  \textbf{Modifiability}.\\
Theories derived from an STGT method application represent the state of practice as a moment in time and are locally generalisable to the studied and similar contexts. In this sense, a credible and rigorously derived STGT theory can never be falsified. A robust theory stands the test of time through its ability to be modifiable in light of new evidences and new contexts. The generalisability of the theory improves as it evolves to accommodate data from varying contexts. Possibilities of deriving strong theories applicable in a wide set of contexts are explored next.

\section{Vision and New Frontiers}
\label{Vision}
This article is an attempt to acknowledge, highlight, and label the unique socio-technical research context of software engineering and introduce Socio-Technical Grounded Theory (STGT). The philosophical, methodological, and evaluation guidelines of STGT presented in this article will be particularly valuable to researchers who are new to theory development, struggling to understand and apply traditional GT methods, unaware or unclear about the different traditional GT versions and how to select one, unsure if they want to conduct a full GT study or only use its data analysis techniques, and unsure how to present and evaluate preliminary, partial, and mature findings of GT studies to a high standard. As shown in Table \ref{tab:TraditionalGTLimits}, if applied well, STGT can help address each of these challenges.

Conversely, STGT may not be applicable in all research contexts. For example, where researchers are interested in studying the social nature of a phenomenon and not in its technical aspects, or where the phenomenon itself is primarily social, the traditional sociological GT methods may be better applicable.

\noindent In summary, the key contribution of this article are to,
\begin{itemize}
    \item Define \textbf{socio-technical research}, with its unique context including -- socio-technical \textit{phenomenon}, \textit{domain and actors}, \textit{researcher}, and \textit{research data, tools, \& techniques} (detailed in section \ref{context} and summarised in Figure \ref{fig:STResearch}).
    \item Present \textbf{Socio-technical Grounded Theory (STGT)} guidelines, including its unique socio-technical research context (as above), philosophical foundations (section \ref{philosophy} and Table \ref{tab:typesofreasoning}), methodological steps and procedures (detailed in sections \ref{method} and \ref{advancedtheorydev}, and summarised in `  \ref{fig:STGT} and Tables \ref{tab:comparison} and \ref{tab:theorydevmodes}), guidelines on application, outcomes, and reporting (detailed in section \ref{application} and captured in Figure \ref{fig:STGTApplications}) and evaluation guidelines (section \ref{evaluation} and Table \ref{tab:evaluation}).
\end{itemize}

\noindent The concepts introduced in this article push research boundaries into new frontiers in many ways.

\begin{itemize}

    
    \item \textbf{\textit{Beyond software engineering}}: STGT has the potential to propel theory development in all contexts where the phenomenon can be best understood by analysing primarily qualitative data and in domains where socio-technical context plays a key role, e.g. SE and related socio-technical disciplines such as information systems, computer science, human computer interaction, artificial intelligence, and user experience research. With increasing proliferation of technology and digitilisation, it may also be useful for classic disciplines such as the social sciences, medicine, and education.
    
    \item \textbf{\textit{Beyond grounded theory}}: As the world becomes increasingly socio-technical, the framework of socio-technical research as defined in this article,  holds relevance beyond STGT, for most research methods such as socio-technical case studies, survey research, and ethnography.
    
    \item \textbf{\textit{Beyond theory development}}: STGT encourages different levels of application, including \textit{full STGT study} for theory development and \textit{STGT for data analysis}, applied in a standalone capacity or in combination with other methods, as part of R\&D studies or large programs.
    
    \item \textbf{\textit{Beyond local theories}}: Grounded theories are widely acknowledged to be local in application. While universality is not an aim or quality criteria of STGT studies, the relevance and generalisability of theories can be improved to cover a greater variety of underlying contexts, up to a point without losing contextual nuances, through the use of modern sources of data (e.g. GitHub, StackOverflow, Twitter), modern approaches to data collection (e.g. mining software repositories) and analysis (natural language processing, sentiment analysis) that make it possible to strengthen, deepen, and expand the scale of grounded theories. 
    
    \item \textbf{\textit{Beyond manual theory development}}: Further advancements in technology and artificial intelligence offers unexplored potential in supplementing, augmenting, and automating parts of qualitative data analysis to ease human effort and improve both the quality and scale of theory development.
    
    \item \textbf{\textit{Beyond human learning}}: Developing theories helps build  collective human knowledge and propel human learning. However, the fundamental systematic and rigorous steps and procedures of the STGT method, i.e. identifying patterns, insights, and relationships through constant comparison across datasets, can also be potentially harnessed as approaches to guide machine and deep learning in AI-based systems.

\end{itemize}

\section*{Acknowledgements}
I sincerely thank Margaret-Anne Storey, Philippe Kruchten, John Grundy, Klaas-Jan Stol, Christoph Treude, Zainab Masood, Steve Adolph, Ingo Mueller, and  Johannes Berglind Söderqvist for providing their thoughtful and invaluable feedback on the drafts. I am also grateful to James Noble and George Allan for their guidance in the early days of my GT journey. I acknowledge the invaluable contributions of Barney Glaser, Anselm Strauss, Juliet Corbin, and Kathy Charmaz to the Grounded Theory research community.

\ifCLASSOPTIONcaptionsoff
  \newpage
\fi

\bibliographystyle{dcu}
\bibliography{Hoda-STGT-TSE2021V2a}

\begin{IEEEbiography}[{\includegraphics[scale=0.25]{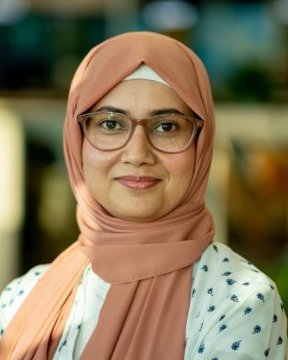}}]{Rashina Hoda} is an Associate Professor of Software Engineering in the Faculty of Information Technology at Monash University, Melbourne. Rashina specialises in use of Grounded Theory in Software Engineering. Rashina received a distinguished paper award for her \textit{grounded theory of becoming agile} at the IEEE International Conference on Software Engineering, ICSE2017. She serves on the Review Board of the IEEE Transactions on Software Engineering and the Advisory Board of the IEEE Software. She presented a Technical Briefing on ``decoding grounded theory for software engineering'' at ICSE2021. Rashina is currently writing a book detailing the \textit{Socio-Technical Grounded Theory} method introduced in this article. For more information visit: www.rashina.com
\end{IEEEbiography}

\end{document}